\documentclass{aa}
\usepackage{graphicx}
\usepackage{wasysym}

\begin{document}

  \title{The non--LTE line formation of neutral carbon\\
                    in late-type stars}

  \author{D. Fabbian\inst{1},
     M. Asplund\inst{1},
     M. Carlsson\inst{2,3},
     D. Kiselman\inst{4} }

  \offprints{D. Fabbian, \\
\email{damian@mso.anu.edu.au} }

  \institute{Research School of Astronomy \& Astrophysics, The Australian
  National University, Mount Stromlo Observatory, Cotter Road, Weston
  ACT 2611, Australia
        \and Institute of Theoretical Astrophysics, University of
        Oslo, P.O. Box 1029, Blindern, N-0315 Oslo, Norway \and 
        Center of Mathematics for Applications, University of Oslo,
  	P.O. Box 1053 Blindern, N-0316 Oslo, Norway \and The
        Institute for Solar Physics of the Royal Swedish Academy of
        Sciences, AlbaNova University Centre, SE-106 91 Stockholm,
        Sweden}

  \date{Received 8 June 2006 / Accepted 27 July 2006}

  \abstract
  {} %
  {We investigate the non--Local Thermodynamic Equilibrium (non--LTE)
  line formation of neutral carbon in late-type stars in order to remove
  some of the potential systematic errors in stellar abundance analyses
  employing C\,{\sc i} features.}
  {The statistical equilibrium code {\small MULTI} was used on a grid
  of plane-parallel 1D {\small MARCS} atmospheric models.}
  {Within the parameter space explored, the high-excitation C\,{\sc i} lines
  are stronger in non--LTE due to the combined effect of line source function drop
  and increased lower level population; the relative importance of the two effects
  depends on the particular combination of \mbox{$T_{\rm{eff}}$}, \mbox{log $g$},
  \mbox{[Fe/H]}\, and \mbox{[C/Fe]} and on the analysed C\,{\sc i} line.
  As a consequence, the non--LTE abundance corrections are negative
  and can be substantially so, for example $\sim -0.4$~dex in halo turn-off
  stars at \mbox{[Fe/H]}$\sim -3$. The magnitude of the non--LTE corrections is
  rather insensitive to whether inelastic H collisions are included or
  not.}
  {Our results have implications on studies of nucleosynthetic
  processes and on Galactic chemical evolution models. When applying
  our calculated corrections to recent observational data, the upturn
  in [C/O] at low metallicity might
  still be present (thus apparently still necessitating contributions
  from massive Pop. III stars for the carbon production), but at a
  lower level and possibly with a rather shallow trend of $\sim -0.2$~dex/dex
  below [O/H]$\sim -1$.}


  \authorrunning{D. Fabbian et al.}
  \titlerunning{C\,{\sc i} non--LTE spectral-line formation in late-type stars}

  \maketitle

\section{Introduction}

Carbon is synthesised through the $3\alpha$ nuclear reaction in
stellar interiors (Burbidge et al. 1957; Arnett 1996; Pagel 1997). It
is one of the most abundant elements in the Universe, together with
hydrogen, helium, oxygen and neon. It can be found in a variety of
compounds, so that it forms the basis of organic chemistry and of life
(at least as we know it). Carbon is also important for dust formation
processes in the interstellar medium (e.g.  Dopita \& Sutherland
2003). Since it has a quite high ionization potential (11.26~eV), it
remains mostly neutral in the photosphere of solar-type stars. Stellar
spectroscopists thus rely on high-excitation permitted C\,{\sc i}
features and on the forbidden [C\,{\sc i}] line at $872.7$~nm, as well
as on various molecular species (mainly CH and C$_2$), to derive solar
and stellar C abundances (e.g. Asplund et al. 2005a).

A number of studies have investigated the carbon content of disk stars
in the Galaxy (e.g. Andersson \& Edvarsson 1994; Gustafsson et al.
1999), arguing that [C/Fe]\footnote{The usual spectroscopic notation
[A/X]$=\log({\rm N_{A}/N_{X} })_{*}-\log({\rm N_{A}/N_{X} })_{\odot}$
is used throughout} is slowly decreasing with time (i.e. with
increasing metallicity) and probably mostly contributed by massive
stars (MSs). In particular, Gustafsson et al. rule out a {\it main}
origin of carbon in low-mass stars while they argue that the
contribution of intermediate-mass stars to the production of carbon in
the Galaxy is unclear. Takeda \& Honda (2005) have recently studied
the CNO abundances of 160 disk F, G, and K dwarfs and subgiants with
$-1<$\mbox{[Fe/H]}$<+0.4$ to look for any differences between the abundances of
stars with and without planets. In particular, when taking their
calculated non--LTE corrections into account, they also found that
[C/Fe] progressively decreases with increasing metallicity for these
galactic disk stars, by around $\sim 0.2$~dex per dex. A similar slope
has been found for [C/Fe] by Reddy et al. (2003, 2006) in their
analysis of a number of elements in a large sample of F and G dwarfs
in the thin- and thick-disk. In contrast with the above authors,
Bensby \& Feltzing (2006) find essentially no trend of [C/Fe] versus
\mbox{[Fe/H]}\, for metal-rich thick- and thin-disk stars and conclude that the carbon
enrichment at such metallicities is mainly due to low- and
intermediate-mass stars (LIMSs), with massive stars still playing the
role of main contributors at lower metallicities in the thick
disk/halo.

The halo stellar population is less well studied. It is still
currently a matter of debate whether for halo stars carbon is produced
primarily in LIMSs (e.g. Chiappini et al. 2003), or in massive stars
(e.g. Akerman et al. 2004). It could well be that contributions from
both of these sources operating on different timescales are
important. Akerman et al. were the first to derive a previously
unrecognized [C/O] upturn at low metallicities. However, they warned
about possible non--LTE effects at low metallicity for the
high-excitation C\,{\sc i}\, lines they employed and cautioned that
their results might be prone to uncertainties due to their assumption
that C\,{\sc i}-based abundances would be affected by the same
non--LTE corrections as for O\,{\sc i}\, (which they estimated for
their sample by interpolating between the values given in Nissen et
al. 2002) and thus that the [C/O] ratio would be unchanged with
respect to LTE. They mentioned the possibility that non--LTE effects
for the carbon and oxygen high-excitation lines might differ by an
amount increasing with decreasing metallicity. At that stage, they
were not able to exclude that the [C/O] values were being
overestimated by $\sim 0.2$~dex under the assumption of LTE. The idea
that the upturn could thus be removed is in fact presented in some
recent work (e.g. Takeda \& Honda 2005; Fabbian et al. 2005; Asplund
2005; Bensby \& Feltzing 2006). Spite et al. (2005) study a sample of
extremely metal-poor giants to investigate the abundances of C, N, O
and Li. They find that in their ``unmixed'' stars (which should
reflect the abundances in the early Galaxy) the [C/Fe] ratio is
remarkably constant (at a level of $\sim +0.2-0.3$~dex) over the
entire range $-4.0<$\mbox{[Fe/H]}$<-2.5$ and that the [C/O] ratio is
close to solar at the lowest metallicities.  According to the studies
of Akerman et al. (2004) and Spite et al. (2005), the upturn in the
[C/O] abundances of halo stars present at low metallicity could signal
substantial early nucleosynthetic production of carbon in massive
Pop. III stars. However it is worth noting that the inclusion of
non--LTE/3D effects plays a substantial role in testing such
hypothesis. For example, 3D effects affecting the C and O abundance
indicators used in the study by Spite et al. (2005) might reach $\sim
-0.6$~dex or more for CH, while they likely remain around $-0.2$~dex
for the [O\,{\sc i}] forbidden oxygen lines (Asplund 2005; see also
Collet et al. 2006). This would then make their resulting [C/O] trend
more or less flat at a level of $\sim -0.6$~dex down to very low [O/H]
ratios, implying that resorting to high yields from massive
zero-metallicity stars might be premature.

On the theoretical side, low-metallicity fast-rotating models of
massive stars (Meynet et al. 2006; see also Chiappini et al. 2006)
predict important CNO yields through large mass loss during the red
supergiant phase via stellar winds caused by efficient mixing and
dredge-up enriching the stellar surface in heavy elements. These
yields, especially if used in conjuction with a top-heavy IMF, would
thus give large carbon abundances at low metallicity without
introducing Pop. III stars. Carigi et al. (2005) have constructed
various Galactic chemical evolution models with different C, N and O
yields. The only models that are able to fit the carbon gradient are
those including C yields that increase with metallicity due to stellar
winds in massive stars and decrease with metallicity due to stellar
winds in LIMSs. They argue that the fraction of carbon in the
interstellar medium due to MSs and LIMSs would thus be dependent on
time (with MSs dominating the C enrichment at early times, while
around $12+\log$(O/H)$\sim 8$ the LIMSs formed in the halo should be
at the end of their evolution and they would start contributing with C
ejecta in a comparable amount to that of MSs, see also Bensby \&
Feltzing 2006) and on galactocentric distance (LIMSs giving a larger
contribution at large r$_{\rm GC}$), with similar contribution from
both sources to the present carbon abundance in the interstellar
medium of the solar vicinity. However, no attempt was made in that
work to homogenize the sample of available observational
data. Moreover, their ``best'' models are still not able to properly
fit the N abundance values for metal rich stars and the C/O values at
intermediate or low metallicity obtained from stellar
observations. Gavil\'{a}n et al. (2005) have recently suggested that a
Galactic Chemical Evolution (GCE) model including yields from Woosley
\& Weaver (1995) for massive stars needs to be integrated with yields from
LIMSs to properly reproduce the wealth of observational data on the
[C/Fe] trend down to \mbox{[Fe/H]}$\sim -2.4$, suggesting that a downward
revision of mass loss by stellar winds from massive stars is necessary
and adding to the increasing evidence that LIMSs are significant
contributors to the Inter Stellar Medium (ISM) carbon
enrichment. Interestingly, Ballero et al. (2006) test several sets of
yields for very massive zero-metallicity stars, using the two-infall
chemical evolution model by Chiappini et al. (1997). They argue that
the contribution of massive Population III stars to the Galactic
evolution of carbon, nitrogen and iron would be negligible even at the
lowest metallicities currently probed and that the existence of such
objects could in any case not be proven or excluded on the basis of
available data on known metal-poor halo stars, which likely formed as
second-generation objects after the explosion of primordial
supernovae.

Absorption systems give the opportunity to measure abundances in
protogalactic structures at high redshift. For example, Becker et
al. (2006) recently detected an overabundance of O\,{\sc i}\,
absorption features in spectra of background high-redshift quasars
(QSOs) towards the line-of-sight of Lyman-$\alpha$ forest clouds,
deriving a mean $<[{\rm C/O}]>= -0.31 \pm 0.09$. Moreover, Pettini
(private communication) and Erni et al. (2006) also find [C/O]$\sim
-0.3$ in metal-poor Damped Lyman-$\alpha$ systems (DLAs) with
[O/H]$\sim -2.5$, in reasonable agreement with the values for Galactic
stars (Akerman et al. 2004). These results are obviously precious in
complementing Galactic studies at low \mbox{[Fe/H]}. In particular,
Erni et al. argue that enrichment from a population of massive (10-50
$M_{\odot}$) Pop. III stars exploding as core-collapse supernovae
(SNeII) and possibly as hyper-energetic (E$>10^{51}$~erg) so-called
hypernovae (HNe) is the most likely scenario, in contrast to models
using non-zero metallicity progenitors or other explosion mechanisms,
such as super-massive (140-260 $M_{\odot}$) pair-instability
supernovae (PISNe) or Type Ia supernovae (SNeIa). Levshakov et
al. (2006) derive a very low abundance of $^{13}$C towards a quasar at
${\rm z}=1.15$. This does not support the enrichment of gas by
fast-rotating, very low-Z massive stars (e.g. Meynet et al. 2006) and
puts a tight bound to the possible amount of contributions from
intermediate-mass AGB stars to the C nucleosynthesis in matter sampled
by the line of sight to this system.

The still open problem of the origin of carbon in the early Universe
constitutes the main motivation for the present non--LTE study. The
quality of stellar spectroscopic observations possible at present,
require us to make corresponding progress in models of stellar
atmospheres and line formation processes in order to remove systematic
errors. The impact on Galactic chemical evolution studies of such
non--LTE calculations promises to be rewarding.

\section{C\,{\sc i} non--LTE line formation}

\subsection{Non--LTE preamble}

Spectral lines are powerful diagnostic tools that provide us with the
capability of ``remotely sensing'' the physical conditions of the
matter composing the particular atmosphere in which they form, via the
analysis of the radiation we receive from it. However, the problem of
how these lines are actually formed when energy is transferred through
the atmosphere by radiation, is extremely complex (Mihalas \& Mihalas
2000; Rutten 2003). The abundances of the chemical elements present in
the atmospheres of stars are therefore usually derived relying on the
simplifying assumption of Local Thermodynamic Equilibrium (LTE) to
derive the atomic level occupation numbers (Gray 1992). The visible
electromagnetic radiation we observe from stars comes from their
photosphere, where radiation-matter interaction can play an important
role. Radiation is in fact absorbed and emitted by atoms and converted
to different wavelengths via the interaction with the matter composing
the atmosphere; viceversa the atomic level populations are influenced
by radiation. The LTE treatment assumes that this complex process is
governed by a single temperature parameter.

However, radiative rates often dominate over collisional ones in the
atmospheric layers where the lines are formed and thus the LTE
approximation can give misleading results. When LTE is not assumed a
priori, the Saha-Boltzmann equation can no longer be used to determine
the level populations from local properties. A large number of
radiative and collisional processes needs to be treated simultaneously
to account for the coupling between the populations of different
atomic levels for the particular chemical element of interest. The
major non--LTE effects in late-type stars have recently been reviewed
by Asplund (2005). The main mechanisms are:

\begin{itemize}

\item {\it UV overionization:} neutral minority stage elements (like e.g.
Fe\,{\sc i} in the Sun) can be affected by overionization triggered by
ultraviolet radiation (bound-free pumping). This is caused by a
J$_\nu$-B$_\nu$ radiation excess at short wavelengths (where J$_\nu$
is the angle-averaged mean intensity and B$_\nu$ is the Planck
function), with the radiation temperature exceeding the local electron
temperature;

\item {\it Resonance line scattering and photon losses:} scattering from deep
layers and photon losses can make the line source function drop below
the local Planck function, resulting in non--LTE line strengthening;

\item {\it Infrared overrecombination:} occurs when J$_\nu$ drops below
B$_\nu$ and causes recombination to high-energy-lying states;

\item {\it Bound-bound pumping:} ultraviolet/optical radiation can
cause pumping of bound-bound transitions, increasing the populations
of upper levels; it can moreover cause overionisation (like in
{B\,{\sc i}} or {Li\,{\sc i}}), since other levels than the one
involved in the transition can be directly influenced; and

\item {\it Photon suction:} caused by a recombination downflow from the
first ionization stage, being for example the cause of the observed
12 $\mu$m emission Mg\,{\sc i} features in the Sun.

\end{itemize}

\subsection{Model atom}
\label{fabs:atom}

Given the complex interaction of the different physical mechanisms in
the line formation process, the accuracy of the atomic data needed as
input is crucial to the final reliability of non--LTE calculations.

Our carbon model atom contains 217 energy levels (207 for C\,{\sc i},
9 for C\,{\sc ii} and one for C\,{\sc iii}) and is constructed taking
into detailed account a total of 650 radiative transitions (453
bound-bound and 197 bound-free). The neutral carbon levels are
included up to an excitation potential of 11.13~eV (i.e. $\sim 0.13$~eV
below the continuum, complete to a principal quantum number of
n=9), the singly ionized levels up to 28.87~eV, while the C\,{\sc iii}
level is at 35.64~eV. All radiative transition probabilities (f-values
and photo-ionization cross-sections) are taken from the Opacity
Project database TOPbase (Cunto et al. 1993; see also, specifically
for carbon, Luo \& Pradhan 1989 and Hibbert et al. 1993). The
necessary data for central wavelengths, radiative broadening and
excitation potential for the atomic energy levels were taken from the
NIST\footnote{http://physics.nist.gov/PhysRefData/} (Martin et
al. 1995; Wiese et al. 1996) database.

Our model atom also includes collisional excitations and ionizations
by electrons and neutral hydrogen atoms. The cross-sections for
forbidden transitions by electron collisions are derived with van
Regemorter's (1962) formula (adopting an oscillator strength
f$=0.01$), while the impact approximation is employed for radiatively
allowed transitions. One of the largest sources of uncertainty in
current non--LTE studies is due to the unavailability of theoretical
and laboratory collisional cross-sections for inelastic collisions
with H\,{\sc i}. Hydrogen collisions in general act as a thermalizing
source, bringing the balance of level populations closer to
LTE. Especially in metal-poor stars, where neutral hydrogen atoms
outnumber free electrons, such collisions should be of importance (e.g.
Asplund 2005). We treated these collisions through the use of Drawin's
formula (1968), as generalized by Steenbock \& Holweger (1984) to the
case of collisions between different particles.

\begin{figure*}[ht]
  \centering
  \includegraphics[width=8cm,height=9cm]{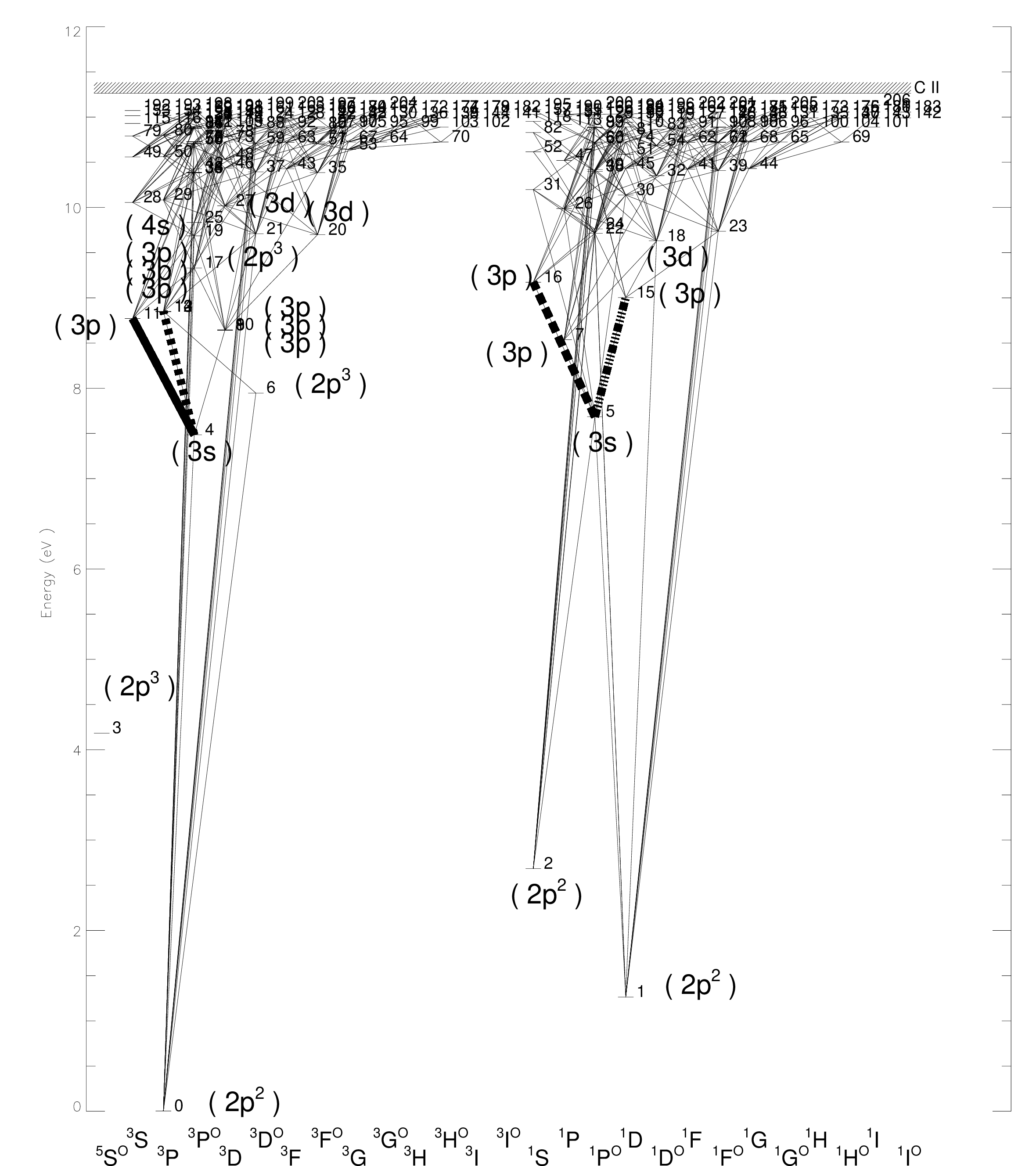}
  \hspace{1cm}
  \includegraphics[width=8cm,height=9cm]{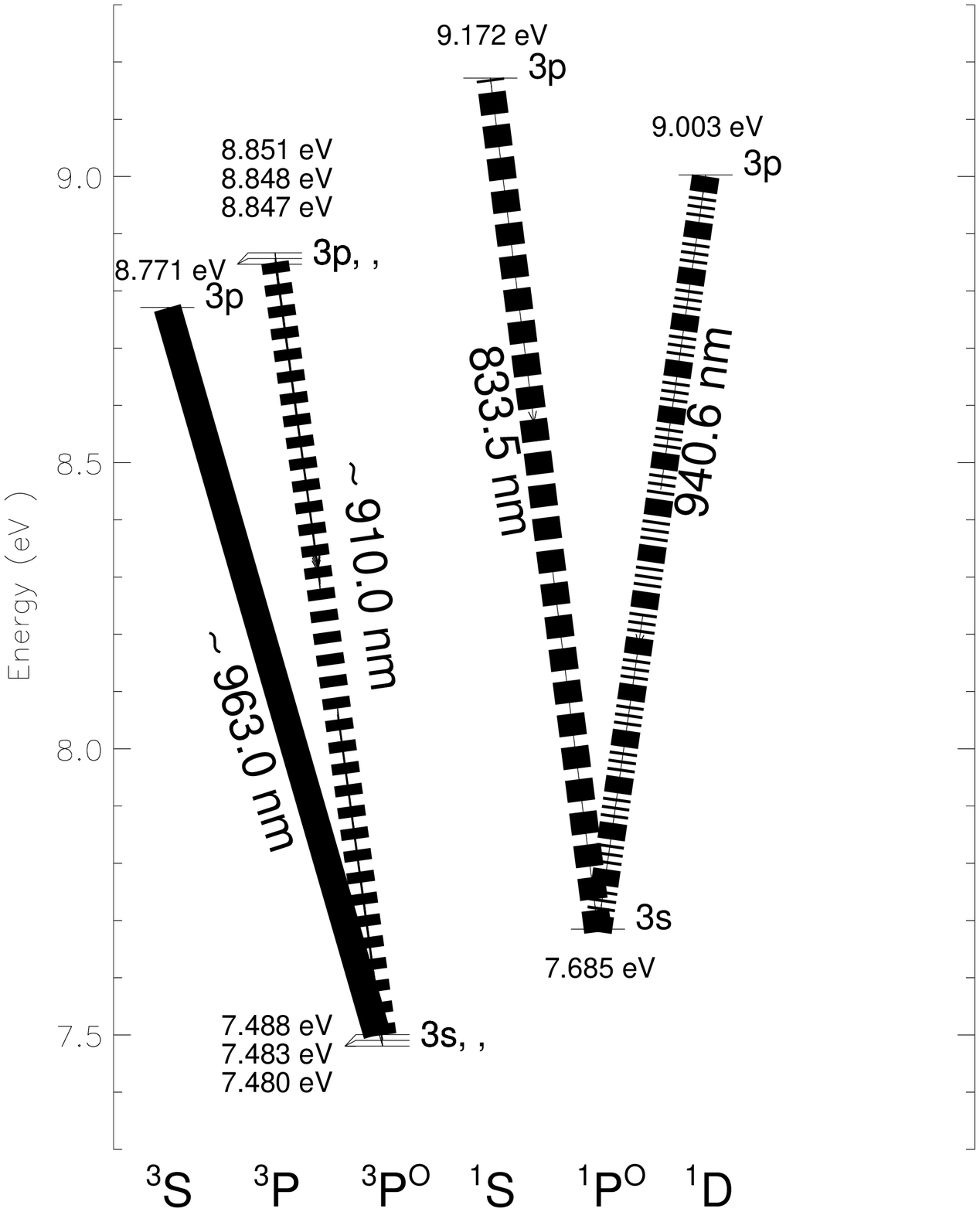}
  \caption{Left panel: Grotrian term diagram for our 217-level (207
  for C\,{\sc i}, 9 for C\,{\sc ii} and one for C\,{\sc iii}) carbon
  model. For clarity, only the 453 bound-bound radiative transitions
  for C\,{\sc i} are shown here. Right panel: Grotrian term diagram
  for the smaller carbon model atom used to compute the non--LTE
  abundance corrections for the C\,{\sc i}\, lines around
  $830.0-960.0$~nm of particular interest here. The model includes 10
  C\,{\sc i}\, levels and 1 C\,{\sc ii} level. The bound-bound
  transitions in the triplet system generate from splitted levels in
  this model, producing three and six lines around $963.0 {\rm ~and~} 910.0$~nm
  respectively}
\label{fabf:termdiag_c}
\end{figure*}

The Grotrian term diagram for our model atom is shown in
Fig.\,\ref{fabf:termdiag_c}. When using this atomic model, we did not
include fine structure splitting for the levels of interest here,
which means that the level populations were computed with compound
lines. We then expect each of these specific ``grouped'' lines to be
stronger (and thus formed in higher atmospheric layers where the
non--LTE effects are likely to be somewhat greater) than the
corresponding observed splitted lines for the same term
transitions. To be able to compare our results with observations, we
have used a second model atom that includes fine splitting,
redistributing the converged population densities according to
statistical weight. Table~\ref{fabt:atom2} gives the relevant details
for the lines included in this smaller atomic model (see also
Fig.\,\ref{fabf:termdiag_c}). Since one could be concerned that
non--LTE effects may be overestimated using this
approach\footnote{Because the fine-structure levels are not directly
included in the whole problem} we have carried out tests which
indicate that only relatively small differences are experienced in
terms of abundance corrections when adopting the more consistent
approach (the results remaining within $\pm 0.05$~dex of our standard
results at all metallicities, the difference becoming even more
negligible towards \mbox{[Fe/H]}$\sim -3$). Since this marginal
difference can be either positive and negative at the different
metallicities, it is unlikely to introduce any systematic bias when
using our computed non--LTE abundance corrections. In these tests, the
radiative and collisional cross-sections in the splitted level case
were reduced accordingly to the increased number of levels with
respect to the unsplitted case, namely redistributing according to
statistical weight, and a smaller 33-level atomic model was used to
speed up the calculations, since we found that only small differences
are obtained with respect to runs using the full 217-level atom (the
results remaining within $\pm 0.05$~dex at all metallicities). This is
indeed a very interesting result and one which we plan to exploit in a
future work on carbon non--LTE line formation using more advanced 3D
atmsopheric simulation snapshots. The 3D non--LTE calculations in fact
become feasible in a reasonable amount of time with this relatively
simple atom.

The standard 217-level atom we use in this study is sufficiently
complex to also include transitions commonly used as solar
diagnostics. For example, our study confirms that the $538.0$~nm
line experiences small non--LTE effects ($< 0.05$~dex in a
solar-like model, as also found by Asplund et al. 2005a): solar
observations could be used to test the non--LTE modelling. This
feature becomes too weak below \mbox{[Fe/H]}$\sim -1$ and thus is
not useful for determining C abundances at low metallicity. In this
work we have instead paid particular attention to the
high-excitation permitted C\,{\sc i} lines around $910.0$~nm, which
are detectable down to low metallicity and have been used by Akerman
et al. (2004). Our aim is to study departures from LTE for these
lines, with a focus on the resulting C abundances in metal-poor halo
stars once the non--LTE corrections are taken into account.

\begin{table}[!t]
 \centering
 \caption{Details of the absorption lines included in the atomic model
 used after converged populations were obtained. A total of 11
 transitions were treated in detail. Wavelength of absorption
 line, upper and lower level identification, excitation potential for lower level
 and oscillator strength are respectively given for each of the transitions included}
 \begin{tabular}{cllcc}
 \hline
 \hline
$\lambda$       & ID$_l$ & ID$_u$ & $\chi_{exc}$ & f$_{ik}$\\
\protect[nm]   &        &        &     [eV]     &         \\
\hline
960.3036 & 3s $^3{\rm P}_0^{\rm o}$ & 3p $^3{\rm S}$ & 7.480 & 0.127\\
962.0781 & 3s $^3{\rm P}_1^{\rm o}$ & 3p $^3{\rm S}$ & 7.483 & 0.120\\
965.8431 & 3s $^3{\rm P}_2^{\rm o}$ & 3p $^3{\rm S}$ & 7.488 & 0.105\\
 \hline	                         		
906.1433 & 3s $^3{\rm P}_1^{\rm o}$ & 2p $^3{\rm P}_2$ & 7.483 & 0.150\\
906.2492 & 3s $^3{\rm P}_0^{\rm o}$ & 2p $^3{\rm P}_1$ & 7.480 & 0.350\\
907.8288 & 3s $^3{\rm P}_1^{\rm o}$ & 2p $^3{\rm P}_1$ & 7.483 & 0.088\\
908.8515 & 3s $^3{\rm P}_1^{\rm o}$ & 2p $^3{\rm P}_0$ & 7.483 & 0.124\\
909.4834 & 3s $^3{\rm P}_2^{\rm o}$ & 2p $^3{\rm P}_2$ & 7.488 & 0.283\\
911.1809 & 3s $^3{\rm P}_2^{\rm o}$ & 2p $^3{\rm P}_1$ & 7.488 & 0.101\\
\hline	                         		
940.5730 & 3s $^1{\rm P}^{\rm o}$ & 3p $^1{\rm D}$ & 7.685 & 0.586\\
\hline	                         		
833.5148 & 3s $^1{\rm P}^{\rm o}$ & 3p $^1{\rm S}$ & 7.685 & 0.110\\
\hline
 \end{tabular}
 \label{fabt:atom2}
\end{table}

\subsection{Method}

The statistical equilibrium code {\small MULTI}\footnote{Available at
http://www.astro.uio.no/$\sim$matsc/mul22/} (Carlsson 1986, 1992) was
used to simultaneously solve the radiative transfer and rate equations
for the C\,{\sc i} non--LTE spectral-line formation.  Complete
redistribution over Voigt line profile has been assumed. A grid of
plane-parallel 1D {\small MARCS} (Gustafsson et al. 1975; Asplund et
al. 1997 and subsequent updates) model atmospheres with varying
\mbox{$T_{\rm{eff}}$}, \mbox{log $g$}\, and \mbox{[Fe/H]}\, has been
employed in the calculations. The atmospheric parameter grid covers
the range of late-type stars, comprising a total of 168 models with
$4500 \le$ \mbox{$T_{\rm{eff}}$} $\le 7000$~K, $2.0 \le$ \mbox{log
$g$}\ $\le 5.0$ [cgs] and $-3 \le$ \mbox{[Fe/H]}\ $\le 0$. In all
cases, unless noted otherwise, the microturbulence velocity has been
assumed to be $\xi=1.00$~km/s. An inherent assumption in our study is
that the departures from LTE for carbon do not influence the given
atmospheric structure. This is reasonable in the sense that carbon is
not an important direct opacity nor a main electron contributor, so
that the atmospheric opacity (and thus the equilibrium structure of
the atmospheric model) will not change significantly even when large
non--LTE corrections are found.

The standard solar C abundance adopted in our non--LTE radiative
transfer calculations has been $\log\epsilon_{C_\odot}=8.40$, very
close to the recently updated value (Asplund et al. 2005a, b).  For
each model atmosphere, the carbon abundance was varied to cover the
range $-0.60 \le$ [C/Fe] $\le +0.60$ in steps of 0.2~dex. We have
repeated all calculations for three different choices of the scaling
factor S$_{\rm{H}}$ for inelastic collisions (S$_{\rm{H}}=0,
10^{-3}\, {\rm and\,} 1$).

\section{Results}

\subsection{Specific stars: Sun, Procyon, HD140283 and G64-12}

Initial tests were carried out for a few models representing specific
stars (Sun, \object{Procyon}, \object{HD 140283}, \object{G64-12}), to
check the effect of changes in different atomic properties and
identify the main driving non--LTE effects for different stellar
parameters.

The high-excitation C\,{\sc i}\, lines studied here are much stronger
in Procyon (\mbox{$T_{\rm{eff}}$}$=6350$~K, \mbox{log $g$}$=3.96$,
\mbox{[Fe/H]}$=0.00$, Allende Prieto et al. 2002) than in the Sun
(5780/4.44/0.00), so that the higher line formation height in Procyon
results in larger non--LTE effects.  In the Sun, non--LTE abundance
corrections vary between $\sim -0.10$ and $\sim -0.25$~dex for the
lines of interest (when neglecting H collisions), while they rise to
between $\sim -0.25$ and $\sim -0.50$~dex in Procyon. In both stars
the drop of the line source function below the Planck function drives
the departure from the LTE approximation.  However, since the level
populations deviate more from the LTE case when moving outwards in the
atmosphere, in the case of Procyon the non--LTE abundance corrections
are more severe.

The low-metallicity of HD 140283 (5690/3.67/-2.40) implies that the
lines are weak and thus formed relatively deep in the atmosphere.
Naively one would then expect the non--LTE effects to be quite
small. Still, the non--LTE abundance corrections in this star vary
between $\sim -0.20$ and $\sim -0.35$~dex, i.e. larger than in the
Sun. This is mainly due to the lower \mbox{log $g$}\, and thus
densities implying less efficient collisions. The lower
\mbox{[Fe/H]}\, also results in smaller electron pressures at a given
optical depth and therefore again less important collisions. The level
populations start to deviate relatively deep in the atmosphere of HD
140283, causing a line source function drop and an increased line
opacity (because of the lower level of the transition getting
overpopulated), both effects acting at the same time to strengthen the
line. The non--LTE line formation in G64-12 (6511/4.39/-3.20/) is
similar to HD 140283: the lower \mbox{log $g$}\, and \mbox{[Fe/H]}\,
and higher
\mbox{$T_{\rm{eff}}$}\, compared to the Sun all work in the same
sense, thus implying even more pronounced non--LTE abundance
corrections, between $-0.2$ and $-0.45$~dex, mainly due to an
overpopulated lower atomic level (opacity effect making the line
stronger than in LTE).

We have investigated the sensitivity of the non--LTE calculations to
the adopted electron and hydrogen collisional cross-sections. We ran
calculations with a few different values for the parameter
S$_{\rm{H}}$ regulating the H collisions efficiency. We found small
abundance variations of the order of $\sim 0.1$~dex when changing
S$_{\rm{H}}$ even by large factors, suggesting that the resulting
non--LTE corrections are not particularly sensitive to the exact
choice. This also guided us in what values to adopt (S$_{\rm{H}}=0$,
$10^{-3}$ and $1$) when successively running full sets of calculations
for the whole grid of stellar models described below. We checked the
effect of electron collisions by artificially decreasing their
cross-sections by a factor of 10. Also in this case, the effects were
small and non-negligible only at low metallicity where they reach
$\sim \pm 0.03$~dex at most.

We finally tested the influence on the non--LTE corrections when
increasing the microturbulence velocity from $1.0$ to $1.5$~km/s. A
higher value for this parameter, introduced in 1D atmospheric models
as an effective (though not perfect) substitute for convective
motions, is expected to affect mostly the stronger saturated lines,
since the atoms will absorb photons at slightly shifted wavelengths
(where the flux is higher) compared to the line center. The stronger
lines will thus be strengthened and lower abundances should be
obtained from them. By appropriately adjusting the microturbulence
parameter, it is possible to remove trends of derived abundances with
excitation potential or line strength which appear when velocity
fields are ignored. As expected, the lines' equivalent widths (both
LTE and non--LTE) increase slightly, by only a few m\AA\, when
increasing the microturbulence to $\xi=1.5$~km/s.  However, the
resulting non--LTE abundance corrections experience insignificant
variations, decreasing by $<0.01$~dex.

\subsection{Non--LTE effects}

\begin{figure*}[!ht]
  \centering
  \includegraphics[width=14cm,height=14cm]{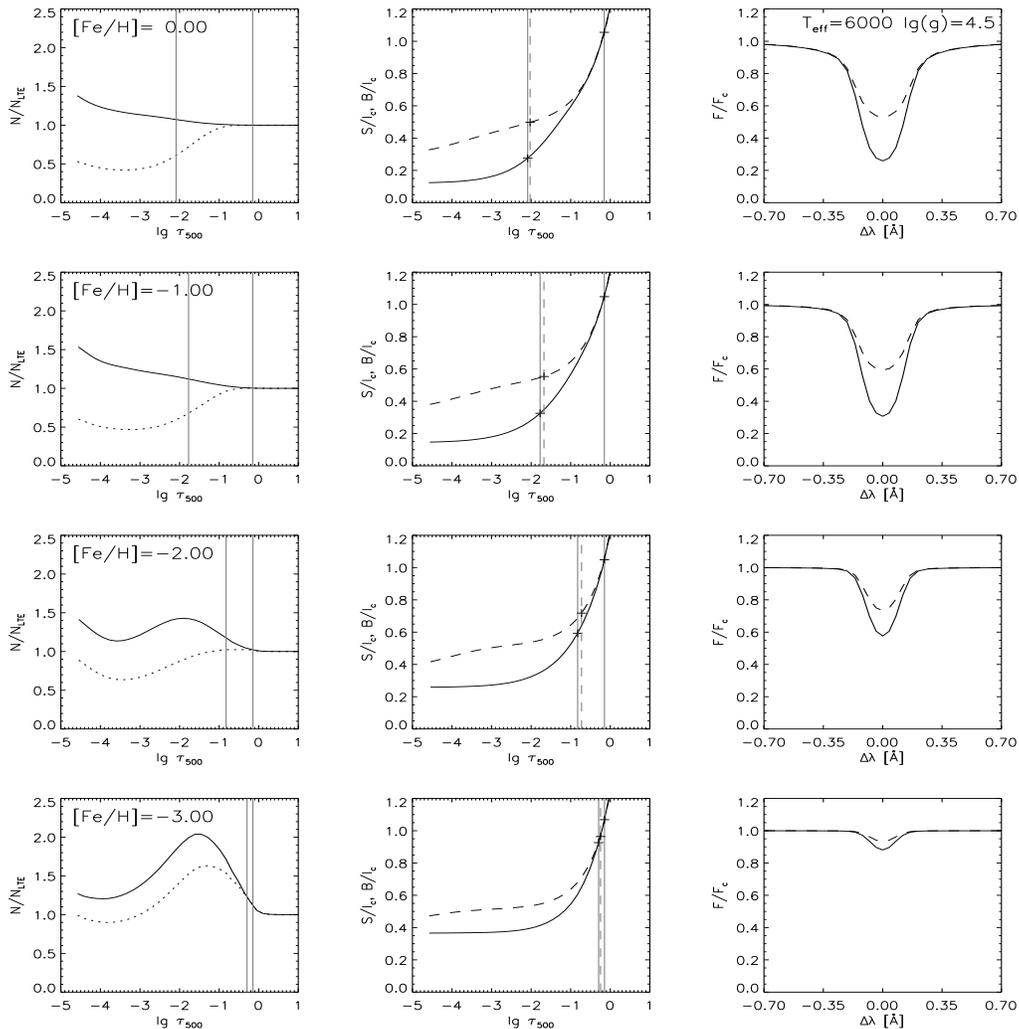}
  \caption{Departure coefficients from LTE (left panels), line source
  and Planck function comparison (middle panels) and resulting line
  profile (right panels) for the C\,{\sc i}\, line at $\sim 910.0$~nm\,
  shown for different \mbox{[Fe/H]}\, for runs with our larger atomic
  model. The atmospheric models all have \mbox{$T_{\rm{eff}}$}$=6000$~K, \mbox{log $g$}$=4.50$
  [cgs] and [C/Fe]$=0$, with \mbox{[Fe/H]}\, going from $0.0$ (top) to $-3.0$
  (bottom). The hydrogen collision efficiency has been set to
  S$_{\rm{H}}=10^{-3}$. For the departure coefficients the solid curve
  represents the lower level and the dotted one the upper level for
  the transition of interest. In the middle panels the line source
  function (solid line) and Planck function (dashed line) are
  normalized to the continuum intensity. Location of
  $\tau_{continuum}=1$ (solid line at {\mbox{log$_{10}~\tau_{500}\sim
  0$}}) and $\tau_{line\, center}=1$ (in LTE and non--LTE, dashed and
  solid line respectively) are shown by vertical lines in the left and
  middle panels. In all cases, the non--LTE line profile (solid line)
  is stronger than in LTE (dashed line)}
\label{fabf:bibetam_t6000g4.50_H0.001}
\end{figure*}

The departure coefficient $\beta$ for an atomic level is defined as
$\beta=N/N_{\rm LTE}$, where N and N$_{\rm LTE}$ are the atomic
populations in non--LTE and LTE respectively, for the level
considered. The ratio of departure coefficients for different atomic
levels enters the line source function\footnote{More explicitly,
assuming complete redistribution and in the Wien regime, the ratio of
the line source function to the Planck function can be expressed in
terms of the ratio of departure coefficients for the upper and lower
levels of a transition, as S$^l_{\nu}$/B$_{\nu}\approx
\beta_u/\beta_l$} S$^l_{\nu}$, so that (when neglecting stimulated
emission, as should be possible for the lines considered here)
identical values of $\beta_u$ and $\beta_l$ (for the lower and upper
level of a transition), result in a Planckian line source
function. This means that even when both levels in a transition have
populations that depart considerably from the Boltzmann expectation,
it is possible that the net effect is a line source function close to
LTE, as long as the departures are the same for the two
levels. However, the effect of line opacity must also be considered
when the lower level population increases or decreases with respect to
the LTE approximation, since it will affect the depth of line
formation.

\begin{flushleft}

\begin{figure*}[!ht]
    \begin{center}
    \includegraphics[width=8.5cm]{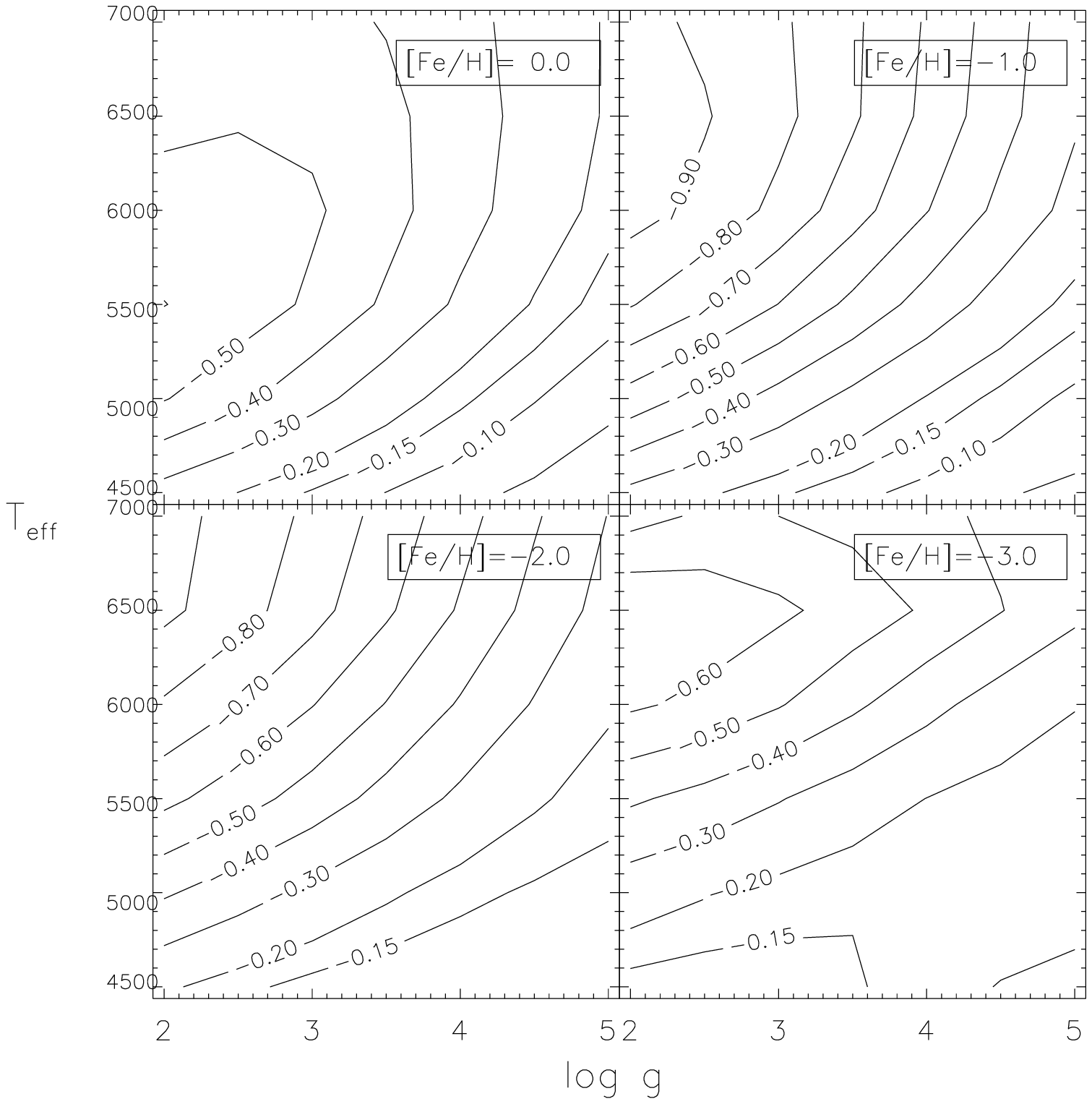}
    \hspace{0.25cm}
    \includegraphics[width=8.5cm]{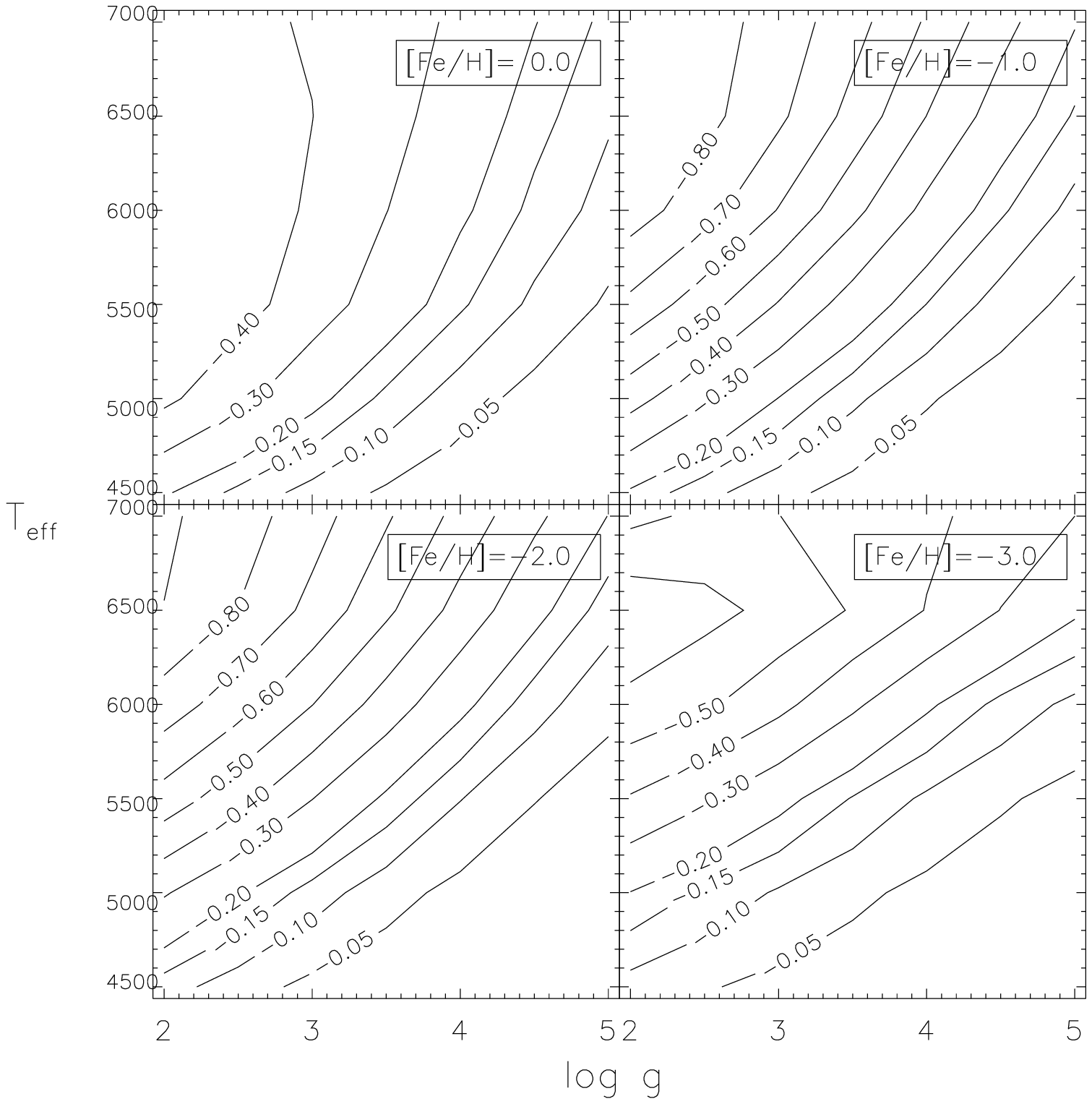}
    \caption{Non--LTE abundance corrections over the grid of
    temperatures and gravities explored, with varying metallicity (as
    indicated in the insets), for the neutral carbon absorption line
    at $909.4834$~nm (used e.g. by Akerman et al. 2004 as one of the
    abundance indicators for their determination of C in halo
    stars). The corrections refer to the case with [C/Fe]$_{\rm
    LTE}=+0.4$, typical of metal-poor halo stars and have been obtained
    from calculations with fine splitting included in the atomic
    model. The four leftmost panels show the corrections we obtained
    when neglecting hydrogen collisions, while the rightmost panels
    represent the results with S$_{\rm{H}}=1$. Large non--LTE
    corrections are present at high temperature and low gravity, in
    the various cases}
\label{fabf:corrcontours}
\end{center}
\end{figure*}

\end{flushleft}

The leftmost panels in Fig.\,\ref{fabf:bibetam_t6000g4.50_H0.001}
show, for a representative case (\mbox{$T_{\rm{eff}}$}$=6000$~K,
\mbox{log $g$}$=4.50$ and collisional efficiency set to
S$_{\rm{H}}=10^{-3}$), the typical departure coefficients $\beta$
for the upper and lower levels of the C\,{\sc i}\, line at $\sim
910.0$~nm. The figure highlights how at solar metallicity the
dominant non--LTE effect is the line source function dropping below
the Planck function, mostly due to the decreased population of the
upper level compared to LTE, while at low metallicity the line
source function is close to LTE, in particular at the much deeper
formation layers in such stars, i.e. closer to the continuum optical
depth; the location of $\tau_{continuum}=1$ and $\tau_{line\,
center}=1$ is marked in the figure. The non--LTE effects are found
to be large even for the most metal-poor stars in our grid. To
explain this, we used the contribution functions defined by Magain
(1986) to evaluate the contribution of the different atmospheric
layers to the formation of the spectral-line depression. Especially
at low metallicity, we found contributions from layers much higher
up than the canonical line optical depth unity to be important. The
non--LTE line strengthening effect can then be explained in such
case, in terms of increased line opacity: the lower level of the
transition gets overpopulated with respect to LTE in those layers,
as seen in the departure coefficients plot for this metallicity
(Fig.\,\ref{fabf:bibetam_t6000g4.50_H0.001}), causing the still
large non--LTE corrections found in this case. In stars with
\mbox{[Fe/H]}$\sim -1$, the two effects just described (line source
function drop and line opacity increase) work in unison, resulting
in large non--LTE effects (see Fig.\,\ref{fabf:corrcontours}).

\begin{flushleft}

\begin{figure*}[!ht]
    \begin{center}
    \includegraphics[width=8.6cm,height=12cm]{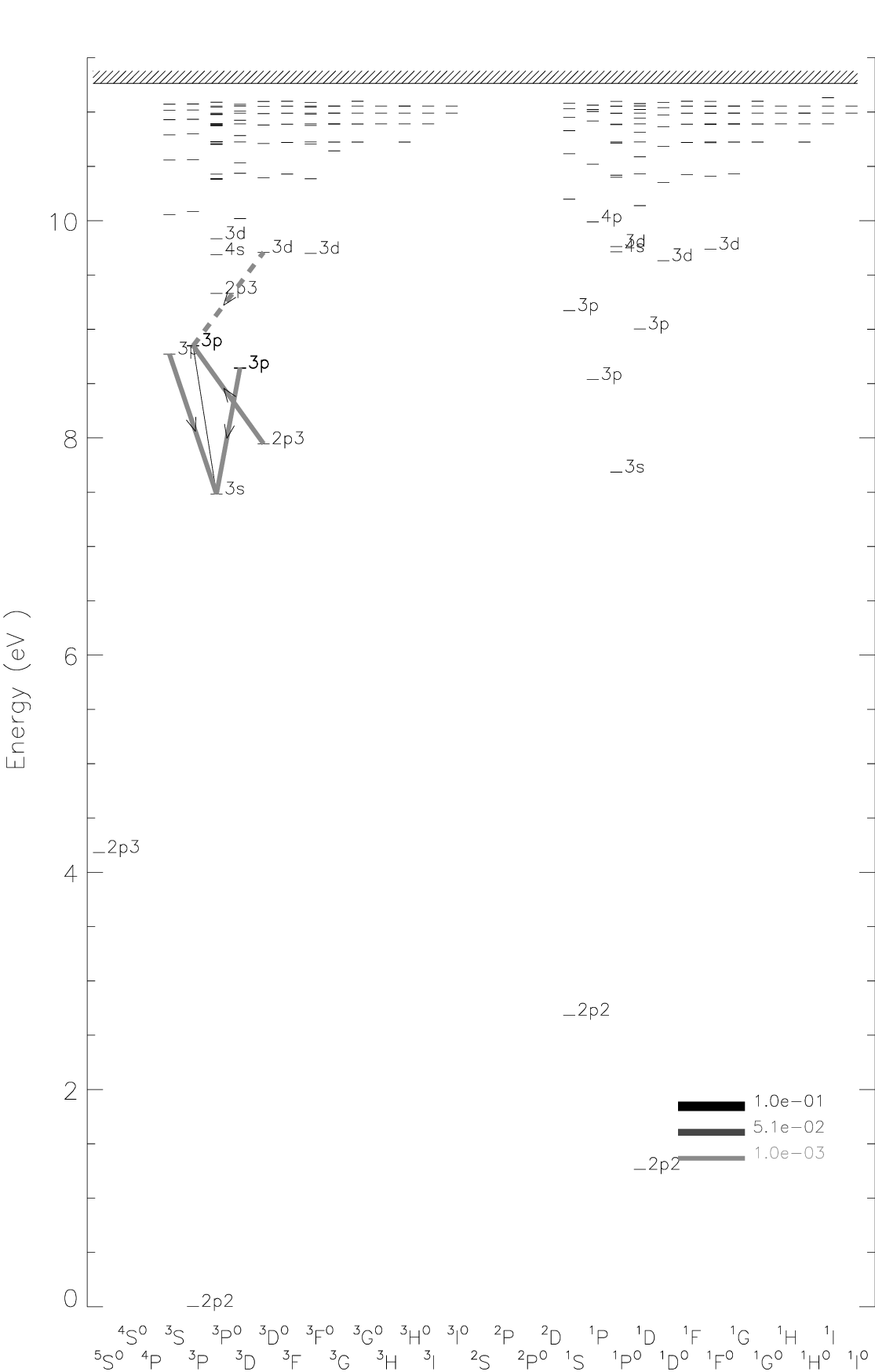}
    \hspace{0.6cm}\includegraphics[width=8.6cm,height=12cm]{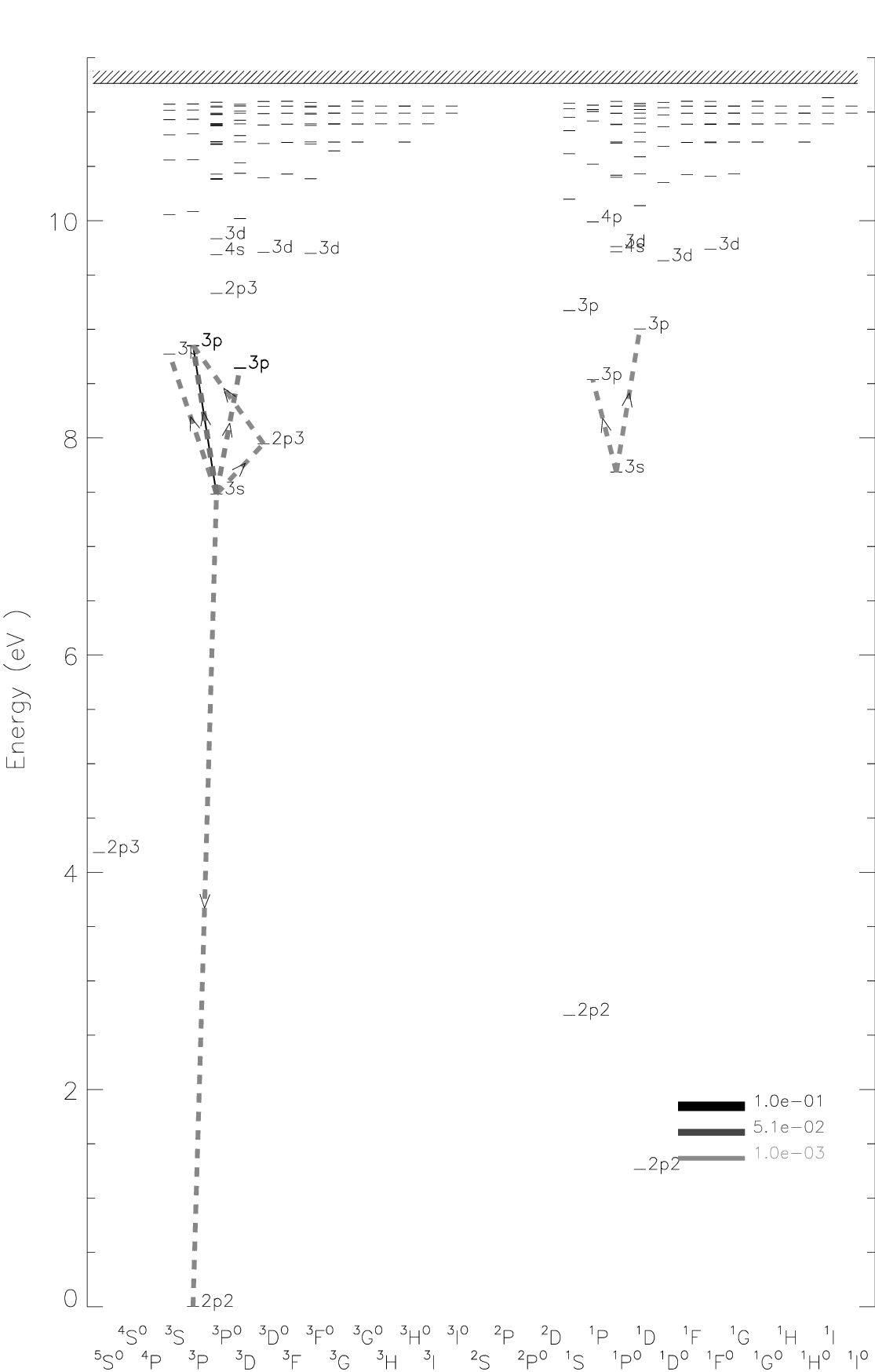}
    \caption{Results from multi-{\small MULTI} runs for a star with
    \mbox{$T_{\rm{eff}}$}$=6000$~K and \mbox{log $g$}$=4.50$ and solar metallicity are shown
    for the C\,{\sc i}\, line around $910.0$~nm. The effect on the line
    strength of multiplying each of the different rates for radiative
    transitions in turn by a factor of 2 is shown in the left
    panel. The right panel shows the effect when varying the
    collisional rates individually by the same amount. Solid lines
    indicate positive contribution (strengthening), dashed lines mean
    negative contribution (weakening) from that particular transition
    to the strength of the line studied.  The thickness indicates the
    relative change in the non--LTE equivalent width (W$^{\rm
    pert}$-W$^{\rm ref}$)/W$^{\rm ref}$ between the perturbed and
    reference case, as in the figure legend, while the arrows indicate
    the net flow (evaluated at $\tau_{line\, center}=1.0$) for the
    transitions}
\label{fabf:multiMULTIsolar}
\end{center}
\end{figure*}

\begin{figure*}[!ht]
    \begin{center}
    \includegraphics[width=8.6cm,height=12cm]{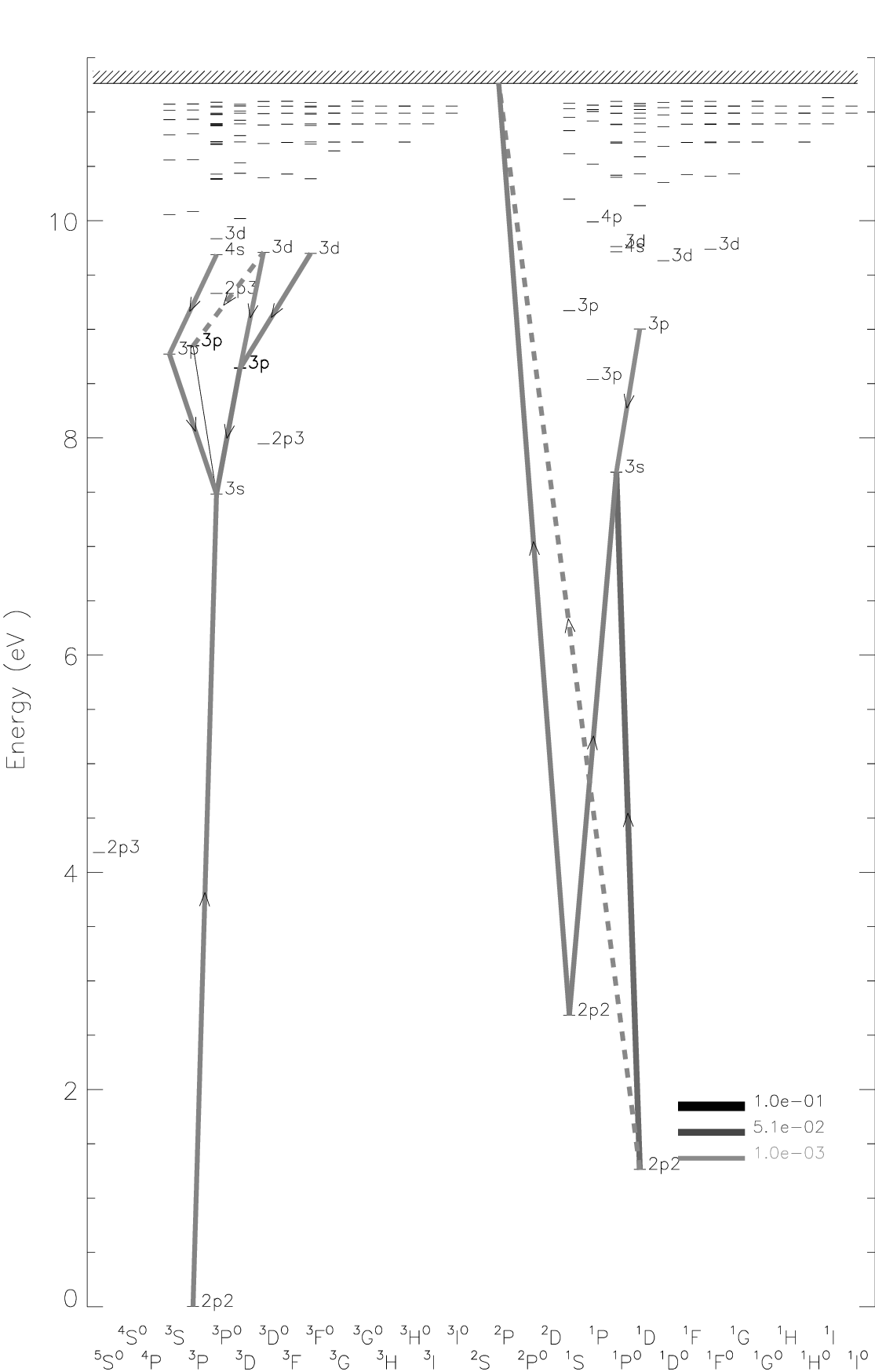}
    \hspace{0.6cm}\includegraphics[width=8.6cm,height=12cm]{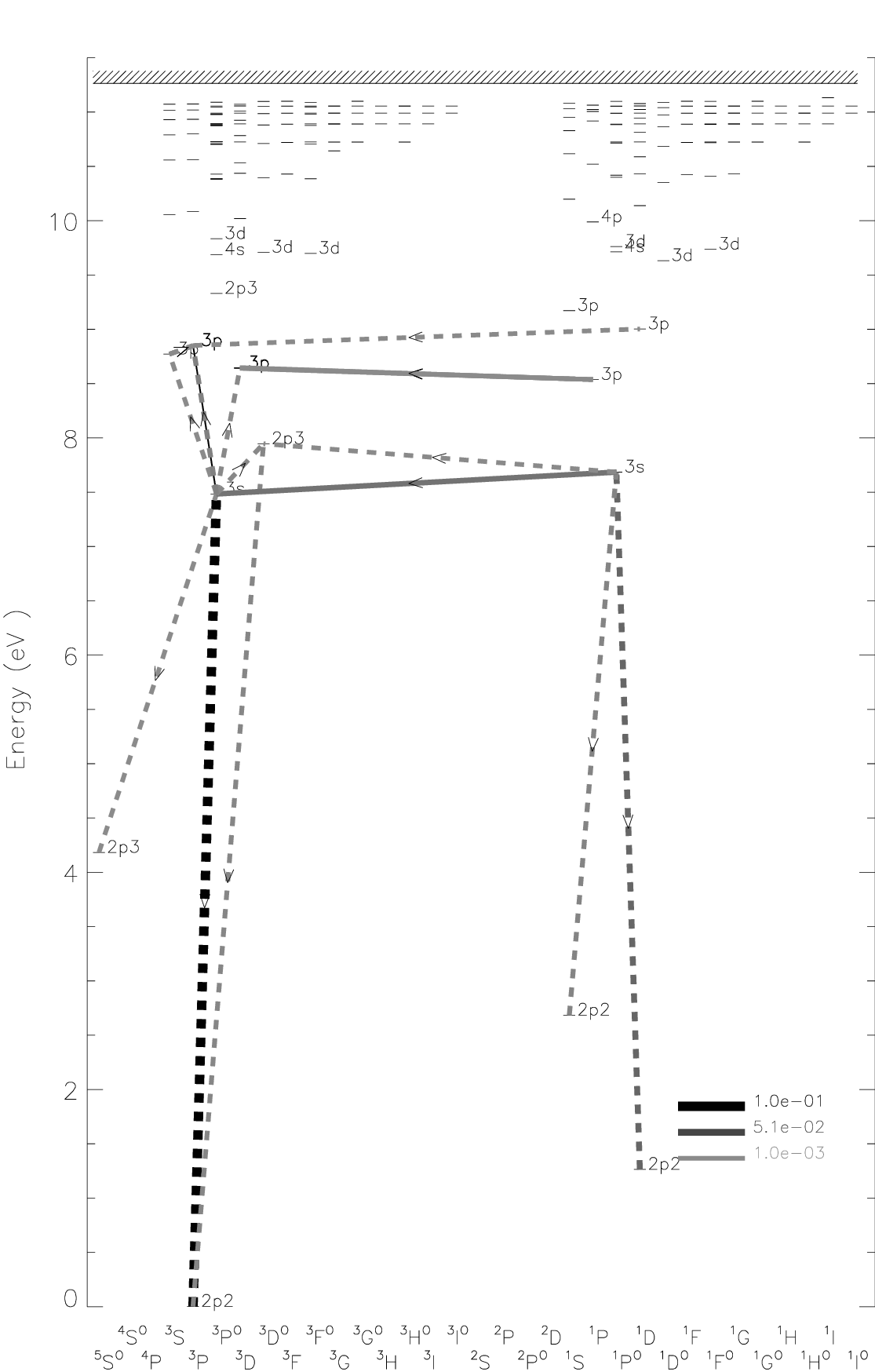}
    \caption{Same as Fig.\,\ref{fabf:multiMULTIsolar}, but for
    \mbox{[Fe/H]}$=-3$}
\label{fabf:multiMULTI-3}
\end{center}
\end{figure*}

\end{flushleft}

Figs.\,\ref{fabf:multiMULTIsolar} and \ref{fabf:multiMULTI-3} show
results from ``multi-{\small MULTI}'' runs. In such calculations, all
650 radiative transitions and a subset of all collisional transitions
(all collisional rates between the lowest 26 atomic levels) were
perturbed in turn by a factor of 2 to study the impact on the C\,{\sc
i}\, lines of main interest here. The runs were carried out for a star
with \mbox{$T_{\rm{eff}}$}$=6000$~K and \mbox{log $g$}$=4.50$ and
metallicities from solar to \mbox{[Fe/H]}$=-3$.  Since for every
transition that is changed in the atomic model a full {\small MULTI}
run has to be performed, the multi-{\small MULTI} runs are very
time-consuming. We refer to Carlsson et al. (1994) for a detailed
description of how such runs are carried out.

\begin{figure*}[!ht]
    \begin{center}
    \includegraphics[width=14cm,angle=90]{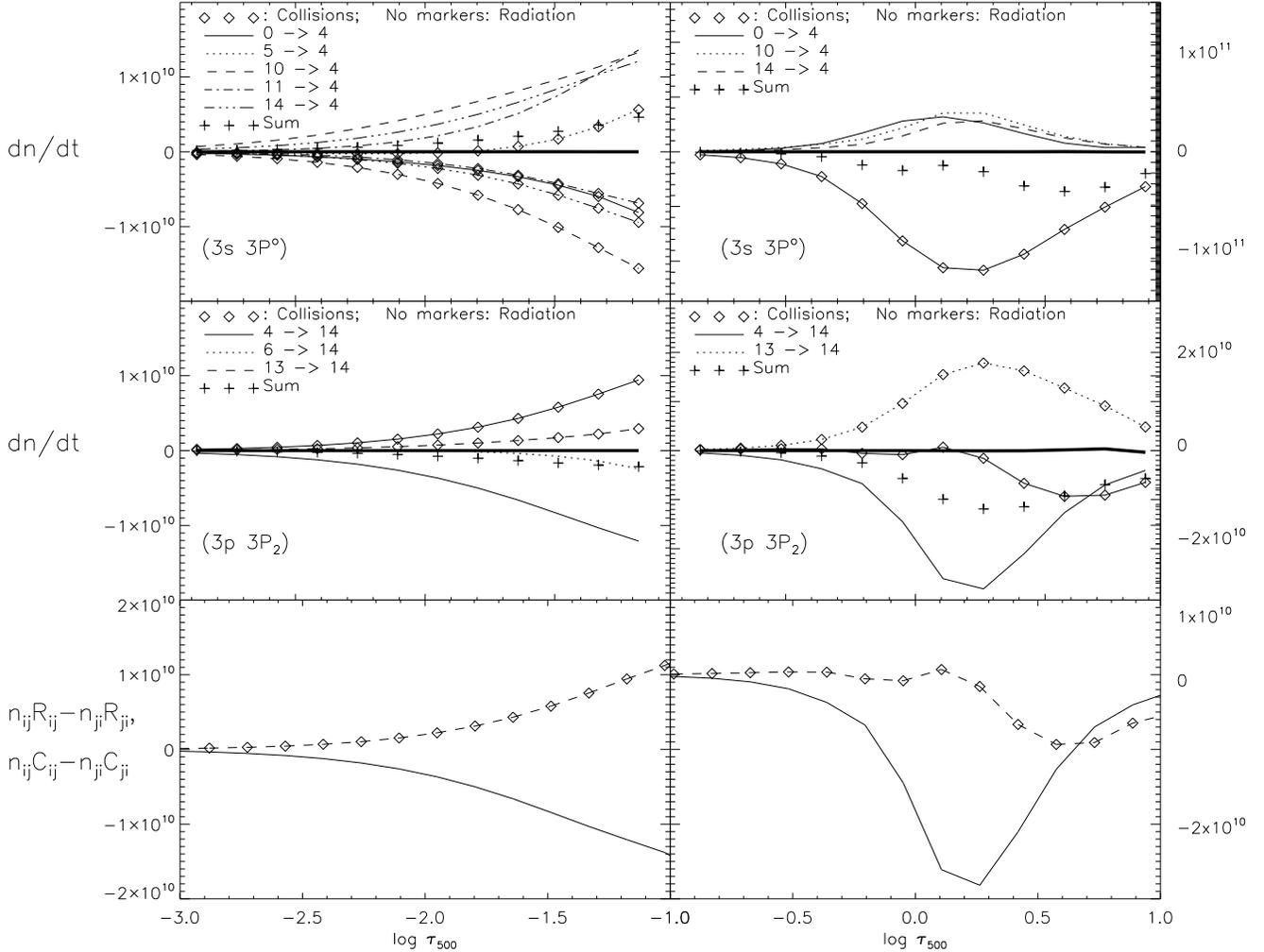}
    \caption{The net rates for radiative (no markers) and collisional
    (with markers) transitions are shown in the upper left plot for
    the lower level of the transition at $\sim 910.0$~nm in a model
    with \mbox{$T_{\rm{eff}}$}$=6000$~K, \mbox{log $g$}$=4.50$ and solar metallicity. Only the
    rates for the transitions important for the level balance are
    shown, marked with different symbols. Their sum is also marked
    with crosses and confirms that the rest of the levels play a minor
    role here. The horizontal line at dn/dt$=0$ marks the expectation
    that the total sum of all rates is equal to zero. The levels in
    the legends are numbered accordingly to our atomic model (see
    Fig.\,\ref{fabf:termdiag_c}), i.e. increasing with
    excitation potential. The middle left plot similarly shows the
    rates for the upper level of the transition, while the bottom left
    plot shows the total sum of the net radiative and collisional
    rates respectively (positive=upwards). The plots on the right show
    the same quantities, but for a model with \mbox{[Fe/H]}$=-3$. Note the
    different scale for the vertical axes in the plots with different
    metallicity}
\label{fabf:levels_balance}
\end{center}
\end{figure*}

Fig.\,\ref{fabf:multiMULTIsolar} confirms that for the $\sim 910.0$~nm
C\,{\sc i}\, feature, the driving non--LTE effect at solar
metallicity is the line source function drop. The $\sim 166.0$~nm
resonance line from the ground state (2p$^2$ 3P - 3s $^3$P$^{\rm o}$)
is an extremely optically thick transition and cannot drain the
population of its upper level (which is the lower level of the $\sim
910.0$~nm transition), which will thus stay close to LTE ($\beta_l
\sim 1$) where the line is formed due to sufficiently strong coupling
to the ground state through electron collisions (see
Fig.\,\ref{fabf:multiMULTIsolar}). The upper transition level instead
gets underpopulated in non--LTE ($\beta_u=N/N_{\rm LTE}<1$), mainly
through photon losses in the line itself, see
Fig.\,\ref{fabf:levels_balance}. Since the ratio $\beta_u/\beta_l$
determines the line source function, this explains why the latter
drops below the Planck function (S$^l_{\nu} < B_{\nu}$) around solar
metallicity (Fig.\,\ref{fabf:bibetam_t6000g4.50_H0.001}), causing the
line strenghtening effect. The effect of collisions (mainly in the
triplet system to which the line studied belongs) is instead to weaken
the line, tending to restore the LTE populations, however the line
source function drop still dominates, driving the non--LTE line
strengthening. At this metallicity, the $\sim 910.0$~nm C\,{\sc i}\,
line will therefore be most sensitive to processes that affect the
upper level population, thus changing the line source function.  There
is very little or no coupling with the singlet system at this
metallicity, so that only a small influence from collisions in the
singlet system is seen. Photoionization is not important either.

For metal-poor stars, the line source function is close to LTE in
the deep layers where the line is formed. Indeed, even though both
levels tend to get overpopulated with respect to LTE, the ratio of
their departure coefficients stays close to the LTE value
($\beta_u/\beta_l\sim1$) where the line is formed\footnote{It is
important to remember that, as mentioned, these weak lines form with
a significant contribution from a tail of atmospheric layers located
much higher up than one would expect in the simple Eddington-Barbier
approach, as we discovered by looking at their contribution
functions (Magain 1986). For example, at
\mbox{$T_{\rm{eff}}$}$=6000$~K, \mbox{log $g$}$4.50$,
\mbox{[Fe/H]}$=-3$, the average height of formation actually moves
out to {\mbox{log$_{10}~\tau_{500}\sim -0.7$}}, with a significant
contribution to the absorption profile still coming from layers as
high up as {\mbox{log$_{10}~\tau_{500}\sim -2$}}. Also note the steep
increase of the level population at low metallicity, as one moves
outward in the atmospheric model
(Fig.\,\ref{fabf:bibetam_t6000g4.50_H0.001})}. However, the onset of
significant departures from LTE in the level population is now shifted
to deeper atmospheric layers than for the solar case.  Moreover, the
overpopulation of the lower level in these layers makes the line form
further out in the atmosphere (where deviations in the level
populations are more dramatic) than in LTE, thus they are stronger
than in that case. This opacity effect causes the line strengthening
to be sensitive to processes affecting the lower level
population. Even though electron collisional coupling to the ground
state still tends to restore LTE population in the lower level of the
$\sim 910.0$~nm transition also at low metallicity, the latter level
now gets severely overpopulated across the atmosphere, due to the fact
that the total sum of the net radiative rates dominates with respect
to collisional ones for this transition (see
Fig.\,\ref{fabf:levels_balance}) and that it cannot depopulate
efficiently and will act as a meta-stable level in detailed balance
with the ground state. Moreover, in the low-metallicity case, the line
strength is influenced by processes in the singlet system much more
than at solar metallicity, i.e. the coupling between the triplet (to
which the lines studied belong) and singlet systems becomes important
(Fig.\,\ref{fabf:multiMULTI-3}) and starts to influence the line
strength more and in a more complex way, both through radiative rates
and collisional transitions. So, while at solar metallicity the
magnitude of the non--LTE effects is mainly governed by collisions and
radiative transitions between levels in the triplet system itself,
with decreasing metallicity both the intersystem (triplet-singlet)
collisions and the processes in the singlet system (collisions as well
as radiation) gain importance with respect to processes in the triplet
system. Electron collisions, especially those between levels with
similar energy, activate the coupling with processes in the singlet
system and thus for example play an important role in determining the
population balance of the lower level of the transition of
interest. The use of accurate collisional rates is therefore
important. Even though at low metallicity the lines studied are weak,
thus formed deeper in than for solar metallicity
(Fig.\,\ref{fabf:bibetam_t6000g4.50_H0.001}), and one could naively
assume the non--LTE effects to be small, it is the relative
strengthening of the line with respect to LTE that justifies the still
large non--LTE corrections we found for low-metallicity stars.  The
main non--LTE effect as one moves to low metallicity is thus an
increased line opacity arising from an overpopulated lower level, with
the contribution coming from the line source function effect still
present but becoming less important with respect to the solar
metallicity, because both the lower and upper levels of the transition
are now similarly overpopulated in the line-forming layers.

Note that in Figs.\,\ref{fabf:multiMULTIsolar} and
\ref{fabf:multiMULTI-3}, the upward or downward direction of the
rates (represented by arrows) is evaluated at $\tau_{line\,
center}=1.0$. In reality such rates can have a complicated behaviour
with atmospheric depth. For example, for the left panel (solar
metallicity, radiative case) in Fig.\,\ref{fabf:multiMULTIsolar},
the rate for the transition from level 2s 2p$^3$ 3D$^{\rm o}$ at
$\sim 7.95$~eV to 2s$^2$ 2p 3p 3P$_2$ at $\sim 8.85$~eV (i.e. from
level 6 to 14, in the notation adopted in
Fig.\,\ref{fabf:termdiag_c}) is upward, but it is found to become
downward just one depth-point below $\tau_{line\, center}=1.0$. The
net result of this is that the upper level population density for
the C\,{\sc i}\, transition around $910.0$~nm actually decreases
when increasing this f-value, which explains why such line becomes
stronger (a line source function effect, because $\beta_u/\beta_l$
becomes less than unity, thus making S$^l_{\nu}<$B$_{\nu}$).

In Fig.\,\ref{fabf:levels_balance} we plot the net radiative and
collisional rates for the transitions that most influence the lower
and upper level of the transition at $\sim 910.0$~nm. One for
example sees that at solar metallicity the population balance of the
upper level (3p $^3{\rm P}_2$, labelled as level 14 in the figure,
as per notation adopted here of increasing level number with
increasing energy, with level 0 being the ground state), shown in
the middle left plot, is set by radiative losses to the lower level
of the transition itself (3s $^3{\rm P}^{\rm o}$) balanced by
collisions from this same level and from level 3p $^3{\rm P}_1$
which has energy close to that of the upper level. The line will
thus get stronger if any of the levels feeding into the upper level
get underpopulated, conversely it will get weaker if we increase the
population of any of these. This picture confirms what is seen in
Fig.\,\ref{fabf:multiMULTIsolar} and described above, namely that
collisions generally have a weakening effect on the line, by tending
to restore LTE populations. The sum of all net radiative rates is
predominant in magnitude with respect to that of collisional rates,
only slightly in the solar-metallicity case, while it strongly
dominates at low metallicity (bottom panels of
Fig.\,\ref{fabf:levels_balance}). The effect is to drive the balance
away from LTE, with a decreased upper level population in the solar
metallicity case mainly due to photon loss in the line itself (the
lower level population remaining close to LTE due to collisional
transitions out of that level which tend to mitigate non--LTE
effects by contrasting the effect of radiative transtions from
higher levels) and an increased population for both levels at low
metallicity (the population flow from higher levels into the lower
level of the transition causing its overpopulation and thus driving
the departure from LTE). As line transitions turn out to be
important, in different ways, both at solar and low metallicity, our
choice of including as many levels and accurate radiative rates in
the atomic model employed (see Sect.\,\ref{fabs:atom}), making it
more complete than those used in some previous studies by other
authors (e.g. St\"{u}renburg \& Holweger 1990; Rentzsch-Holm 1996a,
b), is particularly well suited.

Our results underline that the non--LTE effects for the carbon lines
considered have different causes in different stars. It is thus
paramount to carefully evaluate such effects on a star-to-star basis,
in stellar abundance studies.

\subsection{Non--LTE abundance corrections}

The C\,{\sc i} non--LTE line strength for the parameter grid explored
is generally larger than in LTE. The non--LTE abundance corrections
reach down to as much as $-0.8$~dex and more for particular
combinations of atmospheric parameters (notably at
\mbox{$T_{\rm{eff}}$}$\ge 6000$~K,
\mbox{log $g$}$=2.00$, \mbox{[Fe/H]}$=-1.00$, with corrections of up to $\sim -1.0$~dex).
For each set of atmospheric parameters, the resulting non--LTE
abundance corrections become more severe for stronger lines, similarly
to what is found for the solar case (Asplund et al. 2005a). This can
be explained intuitively because stronger lines will be formed higher
up in the atmosphere where departures from LTE in the level
populations are often more significant.

Several studies (e.g. Kaulakys 1985; Fleck et al. 1991; Belyaev et
al. 1999; Belyaev \& Barklem 2003) now seem to indicate that
Drawin's classical recipe (1968) might overestimate the efficiency
of H collisions by about three orders of magnitude. In view of the
corresponding uncertainty of our final results due to this treatment
of the excitation and ionization by H collisions, we have chosen to
perform full sets of calculations for the whole parameter grid,
applying Drawin's formula with different choices of the scaling
factor S$_{\rm{H}}$ for separate runs, to test how much this affects
our derived non--LTE abundance corrections. We have adopted
S$_{\rm{H}}$=0, 10$^{-3}$ and $1$ respectively and have found a
relatively small sensitivity of the resulting non--LTE corrections
to the value of this parameter. The non--LTE abundance corrections
obtained for all stellar models in our grid, from calculations
adopting the two extreme choices of S$_{\rm H}$ are shown in
Fig.\,\ref{fabf:corrcontours} for the carbon absorption line at
$909.4834$~nm. The plot refers to results obtained when adopting
[C/Fe]$=+0.4$, which is the typical LTE value for the most
metal-poor halo stars in the Akerman et al. (2004) sample. As
expected, we find slightly larger non--LTE abundance corrections
when setting a low efficiency for H collisions. In all cases, stars
in the higher temperature, lower gravity range appear to be
particularly affected, with very large non--LTE effects. The
corrections peak around a metallicity \mbox{[Fe/H]}$\sim -1$. As
seen in the figure, when neglecting inelastic H collisions or
assuming Drawin's recipe (1968) for their treatment, the non--LTE
corrections only vary marginally ($\apprle 0.1$~dex difference
between corrections obtained in the two cases). Since our choices of
S$_{\rm{H}}$ should bracket the true values, it is reassuring that
the corresponding uncertainty in the non--LTE corrections is small.

\begin{figure}[!ht]
  \begin{center}
  \includegraphics[width=9cm,height=6cm]{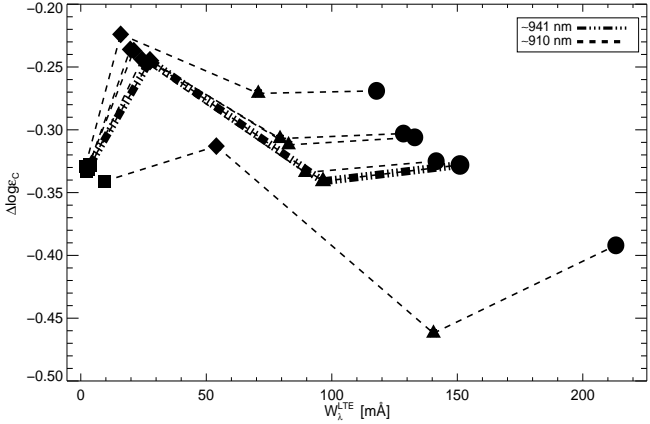}
  \caption{The non--LTE abundance corrections are plotted versus line
  strength for the C\,{\sc i}\, lines around $963.0 {\rm ~and~}
  910.0$~nm. The models have \mbox{$T_{\rm{eff}}$}$=6000$~K\,,
  \mbox{log $g$}$=4.0$ and \mbox{[Fe/H]}$=-3,-2,-1~ {\rm and} ~0$ (for
  which different symbols are used in the figure: filled squares,
  diamonds, triangles and circles respectively). Here [C/Fe]$=0$ and
  S$_{\rm{H}}=10^{-3}$ have been assumed. For each absorption line,
  the non--LTE corrections at different metallicities are connected to
  underline the effect of the metallicity change}
\label{fabf:splittedH0.001}
\end{center}
\end{figure}

At the lowest metallicity, the non--LTE effects are still important,
with $\Delta\log\,\epsilon_{\rm C}\sim -0.35$\, dex for typical halo
stars parameters (\mbox{$T_{\rm{eff}}$} $\simeq 6000$~K, \mbox{log
$g$}$=4.00$) and a choice of S$_{\rm{H}}=10^{-3}$. When we instead
adopt Drawin's recipe as is (parameter regulating collisions with
H\,{\sc i} set to S$_{\rm{H}}$=1), the resulting non--LTE corrections
remain important and are only slightly smaller (still amounting to
$\sim -0.25$~dex at the lowest metallicity for typical halo stars
parameters) than the previous case. Fig.\,\ref{fabf:splittedH0.001}
shows the large non--LTE corrections obtained for a representative set
of atmospheric parameters (\mbox{$T_{\rm{eff}}$}=$6000$~K and
\mbox{log $g$}=$4.00$). They indeed reach down to $\sim -0.45$~dex at
\mbox{[Fe/H]}$=-1$ and are still very substantial
($\Delta\log\,\epsilon_{\rm C}\sim -0.35$~dex) even for weak lines at
\mbox{[Fe/H]}$=-3$. Note that in these plots, only corrections obtained with
[C/Fe]$=+0.00$ are shown. For larger [C/Fe] the line formation is
shifted outwards and hence the non--LTE corrections become even more
severe.

Our analysis aims at removing (some of the) potential systematic
errors in stellar abundance analyses. With this in mind, we provide an
IDL routine\footnote{Available via FTP from:\\
ftp://ftp.mso.anu.edu.au/pub/damian/corr\_nlte/c/} that interpolates
our predicted non--LTE abundance corrections to give the result for
arbitrary stellar parameters and carbon abundances within our grid.

\section{Comparison with previous non--LTE studies}
\label{fabs:comparison}

Only a small number of previous studies have investigated the non--LTE
formation of carbon lines in late-type stars (St\"{u}renburg \&
Holweger 1990; Takeda 1994; Paunzen et al. 1999; Takeda \& Honda 2005;
Asplund et al. 2005a). With the exception of Takeda \& Honda (2005),
these have all been restricted to a few specific stars rather than
covering a large parameter space as we did in our study.

\begin{figure}[!ht]
    \begin{center} \includegraphics[width=8cm]{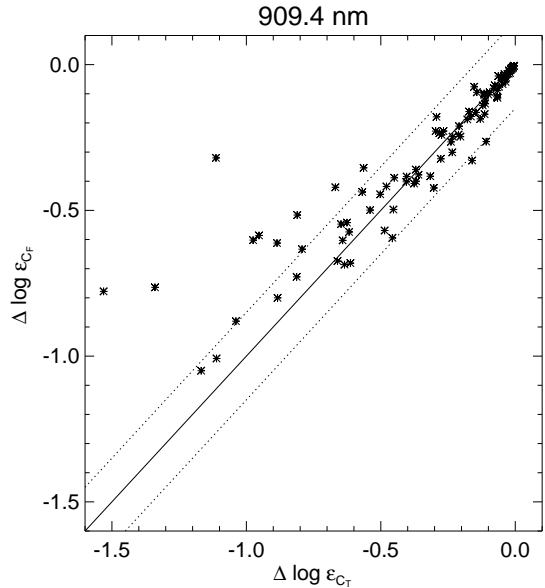}
    \caption{The non--LTE abundance corrections we found in this work
    (vertical axis scale) for the $909.4834$~nm absorption line in a range
    of atmospheric models are compared to the corresponding results in
    Takeda \& Honda (2005), for the case S$_{\rm H}=1$. The solid line
    marks the one to one relation, while the dotted lines include a
    region where the difference between the corrections in the two
    studies is limited to $\le 0.15$~dex}
\label{fabf:Takeda}
\end{center}
\end{figure}

To determine the carbon abundances of their sample of disk stars,
Takeda \& Honda (2005) use the C\,{\sc i} $505.2/538.0$~nm permitted
lines.  These abundance indicators were also used by Asplund et al.
(2005a) for a determination of the solar photospheric carbon
abundance. In both cases, very small non--LTE effects were found for
such lines.  We confirm that these features have small negative
non--LTE abundance corrections ($\Delta\log\epsilon_C$ no larger
than $\sim -0.05$~dex for solar-type stars). To reanalyse the
metal-poor halo stars available in the literature, Takeda \& Honda
carried out non--LTE calculations down to \mbox{[Fe/H]}$=-4$ in
their work, to determine the departure coefficients for the levels
relevant to the same high-excitation features we also study in
detail. They used a carbon model atom with 129 C\,{\sc i}\, levels
(Takeda 1992) for an extensive parameter grid similar to ours but
adopting [C/Fe]$=0$ and $\pm 0.3$ dex only in the calculations. For
the collisional rates due to neutral hydrogen atoms, they also
adopted Drawin's formula in the generalized version by Steenbock \&
Holweger (1984), with the default choice S$_{\rm H}=1$ for all
calculations. We find a generally good agreement between the two
non--LTE studies. However for a few models with high
\mbox{$T_{\rm{eff}}$}\, and low \mbox{log $g$}\, and \mbox{[Fe/H]}\,
our non--LTE abundance corrections, although very large
($\Delta\log\epsilon_C \simeq -0.8$~dex), are not as severe as
claimed by Takeda \& Honda (see our Fig.\,\ref{fabf:Takeda}).

We suspect that the reason for this difference is an extrapolation
error in the work by Takeda \& Honda, possibly due to their use of
only [C/Fe]$=0\, {\rm and\,} \pm 0.3$. For the halo stars available in
the literature, they also find relatively large
($|\Delta\log\epsilon_C| \apprle 0.3$~dex), negative,
metallicity-dependent non--LTE corrections below \mbox{[Fe/H]}$=-2$, those
values however being still $\sim 0.05$~dex smaller than what we obtain
at such low metallicities.

St\"{u}renburg \& Holweger (1990) found that C non--LTE abundance
corrections in the Sun are typically small ($\apprle -0.05$~dex on
average when adopting a scaling factor of 1/3 to Drawin's formula for
the H collisions). For the lines we also study here, they explained
the non--LTE effects in the solar photosphere mainly in terms of lower
level overpopulation, causing the line absorption coefficient to
exceed the LTE value. We underline that the driving effect at solar
metallicity is instead that of the line source function dropping below
the Planck function, i.e. both the lower and upper levels of the
transition need to be considered. The upper level underpopulation is
actually more significant than the lower level (slight) overpopulation
in the line-forming region, driving the line source function drop in
Sun-like stars (see our
Fig.\,\ref{fabf:bibetam_t6000g4.50_H0.001}). In any case, a detailed
comparison between the results for the high-excitation C\,{\sc i}\, lines of
interest shows a very good agreement, the non--LTE abundance
corrections amounting to $\sim -0.1$~dex for those features in both
studies.

Using the same model atom as St\"{u}renburg \& Holweger, Tomkin et
al. (1992) find comparably small non--LTE effects for the C\,{\sc i}\,
$\sim 910.0$~nm lines ($\Delta\log\epsilon_C \apprle -0.15$~dex) also
in metal-poor halo stars. However, they adopted very large
efficiencies for the H collisions, in order to reproduce the observed
line strengths using the high solar carbon abundance expected at the
time. As explained, evidence is now accumulating that such high values
for H collisions might be unrealistic by several orders of magnitude,
so that the non--LTE corrections for the high-excitation C\,{\sc i}\,
features of interest should be expected to be in reality more
significant than in that study.

Using an extended version of the atomic model by St\"{u}renburg \&
Holweger, Rentzsch-Holm (1996a, b) investigated carbon non--LTE
abundance corrections in A-type and related stars. A detailed
comparison with that work is unfortunately not possible, since the
two calculations only overlap for our highest temperature value
($7000$~K). However, for that common temperature value, a good
agreement is found when comparing the resulting non--LTE corrections
for the high-excitation C\,{\sc i}\, lines; at \mbox{log $g$}$=3.5$,
$\Delta\log\epsilon_C$ reaches $\sim -0.45$~dex in both studies. For
a given stellar model, we too find increasing non--LTE corrections
with line strength, see the trend indicated by identical symbols in
Fig.\,\ref{fabf:splittedH0.001}: this effect has already been
explained by St\"{u}renburg \& Holweger (1990) as due to the
numerical enhancement, in the flat part of the curve-of-growth, of
the abundance correction derived from equivalent width changes due
to non--LTE, compared to the abundance correction derived in the
linear part of the curve-of-growth. When decreasing gravity or
increasing effective temperature (see our
Fig.\,\ref{fabf:corrcontours}, note that for
\mbox{$T_{\rm{eff}}$}$\apprge 5500$~K, the non--LTE abundance
corrections tend to become more insensitive to temperature changes
when fixing metallicity and gravity, except at very low
metallicity), generally larger non--LTE corrections result for both
studies. Another similarity between these two works - and also with
that of St\"{u}renburg \& Holweger 1990 - is that the four lowest
levels of C\,{\sc i} (not involved in the high-excitation
transitions of interest here) in our model are found for almost all
cases in the parameter space we study to have Boltzmann populations
throughout the model atmosphere, due to strong collisional coupling
between the neutral ground state and the next three lowest-lying
levels. This ensures that the forbidden [C\,{\sc i}] line at
$872.7$~nm (transition between the two lowest-lying levels in the
singlet system) is in general not affected by departures from LTE.
The levels involved in this transition only start to experience
non--LTE effects at \mbox{[Fe/H]}$=-3$, for
\mbox{$T_{\rm{eff}}$}$\apprge 6500$~K and higher (and increasing
with decreasing gravity too). The two levels tend to experience
rather similar departures from LTE. Rentzsch-Holm (1996b) similarly
found that increasing \mbox{$T_{\rm{eff}}$}\, for the higher
temperature and metallicities range in that study resulted in
underpopulation of these levels: they showed this can transmit to
other levels through line transitions and lead to global
underpopulation of C\,{\sc i} around $\sim 12000$~K. In our case,
both the high temperature and low metallicity act to increase the
magnitude and the atmospheric depth of the onset of departures from
LTE, because the UV flux can more effectively underpopulate the
atomic levels involved in the transition.

\section{Galactic chemical evolution of carbon}
\label{fabs:GCE}

\begin{table*}[!ht]
\begin{center}
 \caption{The IDs, atmospheric parameters (effective temperature,
 gravity and iron content) and the LTE abundances for oxygen and
 carbon for the sample of low-metallicity stars from Akerman et
 al. (2004) are shown in column 1, 2, 3, 4, 5 and 7 respectively. The
 corresponding LTE [C/O] values are in column 10. Our adopted non--LTE
 abundance corrections for oxygen and carbon (obtained when neglecting
 collisions with H\,{\sc i} atoms) and the resulting [O/H] and [C/O]
 non--LTE abundance ratios from this study are shown in column 6, 8, 9
 and 11 respectively. All values from the Akerman et al. study have
 been corrected, where appropriate, for our different choice of the
 carbon and oxygen solar abundances ($\log\epsilon_{C_\odot}=8.39$
 and $\log\epsilon_{O_\odot}=8.66$)}
\begin{footnotesize}
\begin{tabular}{l|ccc|cccccccc}
\hline \hline
Star & \mbox{$T_{\rm{eff}}$} & \mbox{log $g$} & \mbox{[Fe/H]} & $\log\epsilon_O$ & $\Delta\log\epsilon_O$ & $\log\epsilon_C$ & $\Delta\log\epsilon_C$ & [O/H] & [C/O] & [C/O]\\
            &  [K]  & [cgs] & (LTE)&     (LTE)        &                          &      (LTE)       &                  & (non--LTE)    & (LTE) & (non--LTE)\\
\hline
BD-13$^{\circ}$3442  & 6500 & 4.16 & -2.61 & 6.85 &  -0.18*& 6.13 & -0.37 &  -1.99*& -0.45 &  -0.64*\\
CD-30$^{\circ}$18140 & 6272 & 4.13 & -1.88 & 7.55 & -0.13 & 6.72 & -0.26 & -1.24 & -0.56 & -0.69\\
CD-35$^{\circ}$14849 & 6125 & 4.11 & -2.41 & 7.12 & -0.11 & 6.40 & -0.30 & -1.65 & -0.45 & -0.64\\
CD-42$^{\circ}$14278 & 5812 & 4.25 & -2.12 & 7.42 & -0.13 & 6.62 & -0.22 & -1.37 & -0.53 & -0.62\\
HD103723    	     & 6040 & 4.26 & -0.82 & 8.34 & -0.20 & 7.63 & -0.23 & -0.52 & -0.44 & -0.47\\
HD105004    	     & 5919 & 4.36 & -0.86 & 8.26 & -0.19 & 7.75 & -0.22 & -0.59 & -0.24 & -0.27\\
HD106038    	     & 5919 & 4.30 & -1.42 & 8.08 & -0.18 & 7.40 & -0.27 & -0.76 & -0.41 & -0.50\\
HD108177    	     & 6034 & 4.25 & -1.74 & 7.80 & -0.15 & 6.94 & -0.24 & -1.01 & -0.59 & -0.68\\
HD110621    	     & 5989 & 3.99 & -1.66 & 7.95 & -0.17 & 7.06 & -0.28 & -0.88 & -0.62 & -0.73\\
HD121004    	     & 5595 & 4.31 & -0.77 & 8.71 & -0.23 & 8.05 & -0.20 & -0.18 & -0.39 & -0.36\\
HD140283    	     & 5690 & 3.69 & -2.42 & 7.11 & -0.11 & 6.34 & -0.26 & -1.66 & -0.50 & -0.65\\
HD146296    	     & 5671 & 4.17 & -0.74 & 8.53 & -0.22 & 7.96 & -0.24 & -0.35 & -0.30 & -0.32\\
HD148816    	     & 5823 & 4.14 & -0.73 & 8.65 & -0.22 & 8.02 & -0.25 & -0.23 & -0.36 & -0.39\\
HD160617    	     & 5931 & 3.77 & -1.79 & 7.42 & -0.13 & 6.74 & -0.28 & -1.37 & -0.41 & -0.56\\
HD179626    	     & 5699 & 3.92 & -1.14 & 8.46 & -0.22 & 7.62 & -0.25 & -0.42 & -0.57 & -0.60\\
HD181743    	     & 5863 & 4.32 & -1.93 & 7.66 & -0.14 & 6.82 & -0.21 & -1.14 & -0.57 & -0.64\\
HD188031    	     & 6054 & 4.03 & -1.79 & 7.75 & -0.15 & 6.87 & -0.26 & -1.06 & -0.61 & -0.72\\
HD193901    	     & 5672 & 4.38 & -1.12 & 8.16 & -0.18 & 7.41 & -0.20 & -0.68 & -0.48 & -0.50\\
HD194598    	     & 5906 & 4.25 & -1.17 & 8.19 & -0.19 & 7.48 & -0.23 & -0.66 & -0.44 & -0.48\\
HD215801    	     & 6005 & 3.81 & -2.29 & 7.22 & -0.12 & 6.34 & -0.30 & -1.56 & -0.61 & -0.79\\
LP815-43    	     & 6533 & 4.25 & -2.67 & 6.61 &  -0.25*& 6.14 & -0.39 &  -2.30*& -0.20 &  -0.34*\\
G011-044    	     & 5995 & 4.29 & -2.09 & 7.47 & -0.13 & 6.63 & -0.22 & -1.32 & -0.57 & -0.66\\
G013-009    	     & 6360 & 4.01 & -2.27 & 7.15 & -0.11 & 6.47 & -0.35 & -1.62 & -0.41 & -0.65\\
G016-013    	     & 5602 & 4.17 & -0.76 & 8.75$^{\dag}$ & -0.20 & 8.00$^{\dag}$ & -0.22 & -0.11$^{\dag}$ & -0.48 & -0.50$^{\dag}$\\
G018-039    	     & 5910 & 4.09 & -1.52 & 8.12 & -0.18 & 7.29 & -0.25 & -0.72 & -0.56 & -0.63\\
G020-008    	     & 5855 & 4.16 & -2.28 & 7.53 & -0.13 & 6.66 & -0.24 & -1.26 & -0.60 & -0.71\\
G024-003    	     & 5910 & 4.16 & -1.67 & 7.62 & -0.14 & 6.71 & -0.22 & -1.18 & -0.64 & -0.72\\
G029-023    	     & 5966 & 3.82 & -1.80 & 7.79 & -0.15 & 6.85 & -0.28 & -1.02 & -0.67 & -0.80\\
G053-041    	     & 5829 & 4.15 & -1.34 & 7.86 & -0.16 & 7.07 & -0.25 & -0.96 & -0.52 & -0.61\\
G064-012    	     & 6511 & 4.39 & -3.17 & 6.44 &  -0.30*& 5.68 & -0.45 &  -2.52*& -0.49 &  -0.64*\\
G064-037    	     & 6318 & 4.16 & -3.12 & 6.44 &  -0.30*& 5.72 & -0.42 &  -2.52*& -0.45 &  -0.57*\\
G066-030    	     & 6346 & 4.24 & -1.52 & 7.93 & -0.13 & 6.93 & -0.27 & -0.89 & -0.73 & -0.87\\
G126-062    	     & 5943 & 3.97 & -1.64 & 7.98 & -0.17 & 7.02 & -0.27 & -0.85 & -0.69 & -0.79\\
G186-026    	     & 6273 & 4.25 & -2.62 & 6.71 &  -0.15*& 6.14 & -0.33 &  -2.10*& -0.30 &  -0.48*\\
\hline
\end{tabular}
\end{footnotesize}
\end{center}

*Note that the O\,{\sc i}\, non--LTE abundance corrections for the five most
metal-poor stars have been revised with respect to those given in
Akerman et al. (2004), as explained in the text\\

$^{\dag}$ The C and O abundances of G016-013 have been revised by
$+0.4$~dex with respect to those in Akerman et al. (2004), due to
the fact that the wrong (hotter) stellar model for HD160617 was
accidentally used in that analysis instead of the correct one for
this star (P.E.  Nissen, private communication). This affects [O/H]
(and [C/H]) but leaves [C/O] intact for this star. We trust that the
stellar parameters for this star in Akerman et al. (also given here)
have been correctly derived in their analysis

\label{fabt:sample}
\end{table*}

\begin{flushleft}

\begin{figure*}[!ht]
    \begin{center}
    \includegraphics[width=14cm]{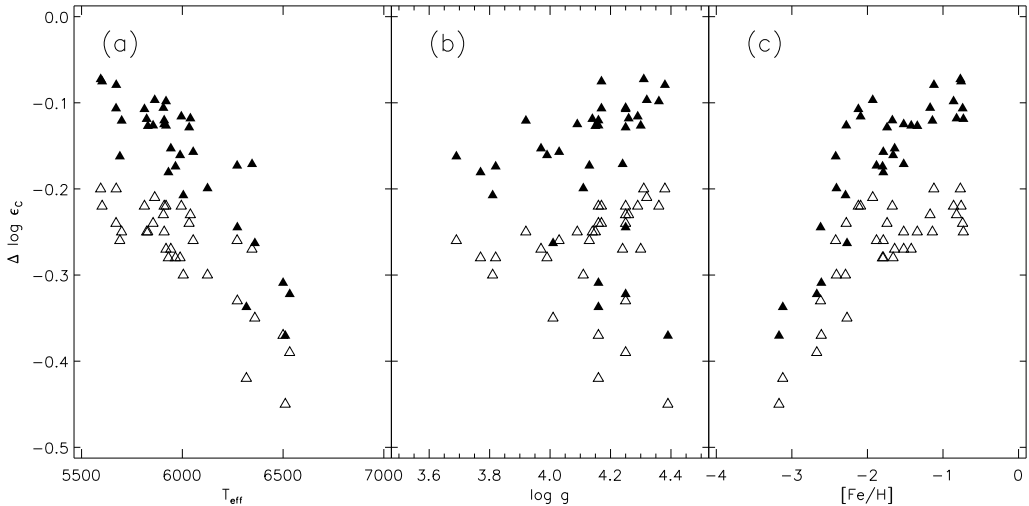}
    \caption{The non--LTE abundance corrections (empty triangles:
    S$_H=0$; filled triangles: S$_H=1$) for the 34 halo stars in the
    sample by Akerman et al. (2004) as a function of \mbox{$T_{\rm{eff}}$}\, (a),
    \mbox{log $g$}\, (b) and \mbox{[Fe/H]}\, (c)}
\label{fabf:corrteff}
\end{center}
\end{figure*}

\end{flushleft}

\begin{flushleft}

\begin{figure*}[!ht]
\begin{center}
\includegraphics[width=8cm]{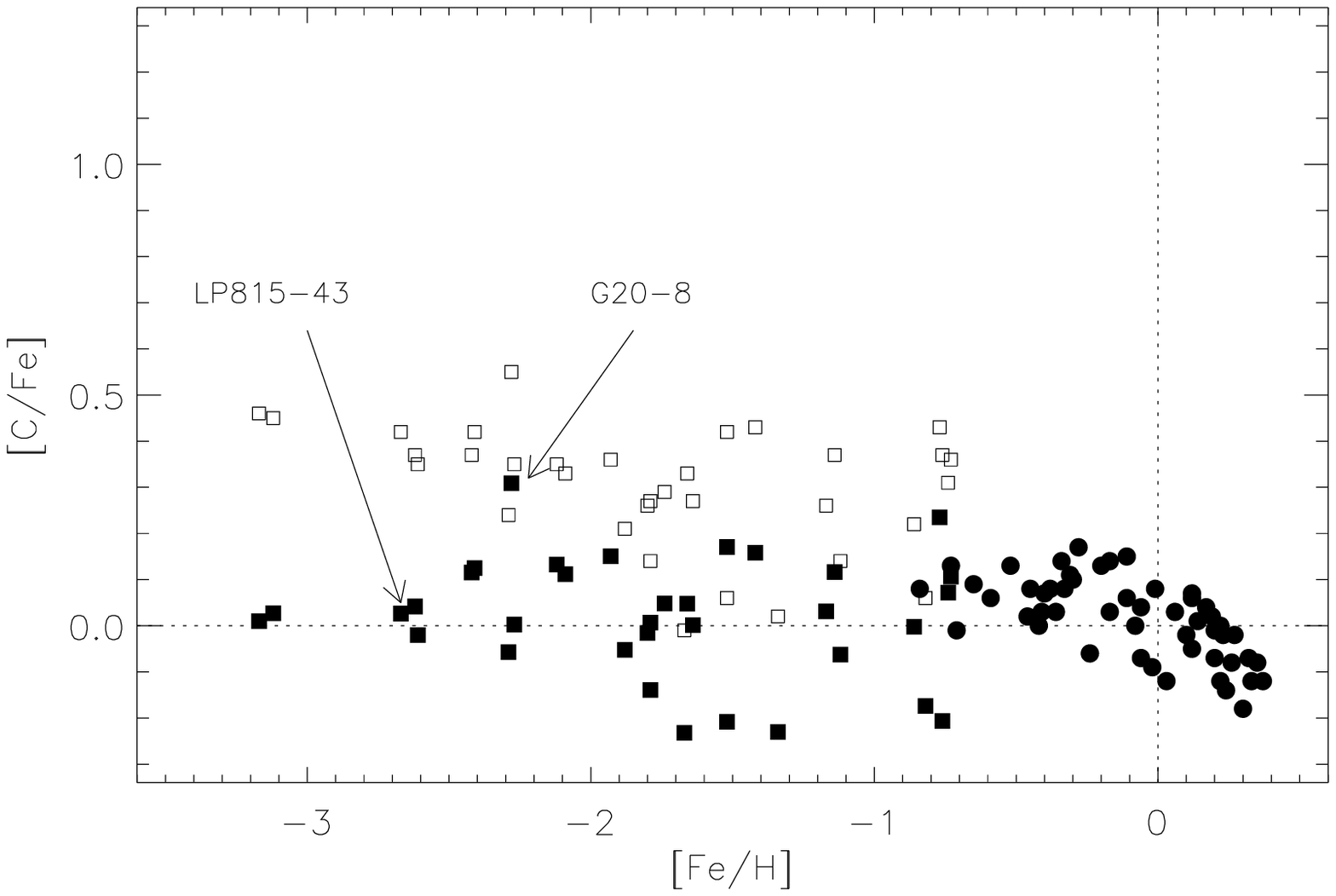}
\includegraphics[width=8cm]{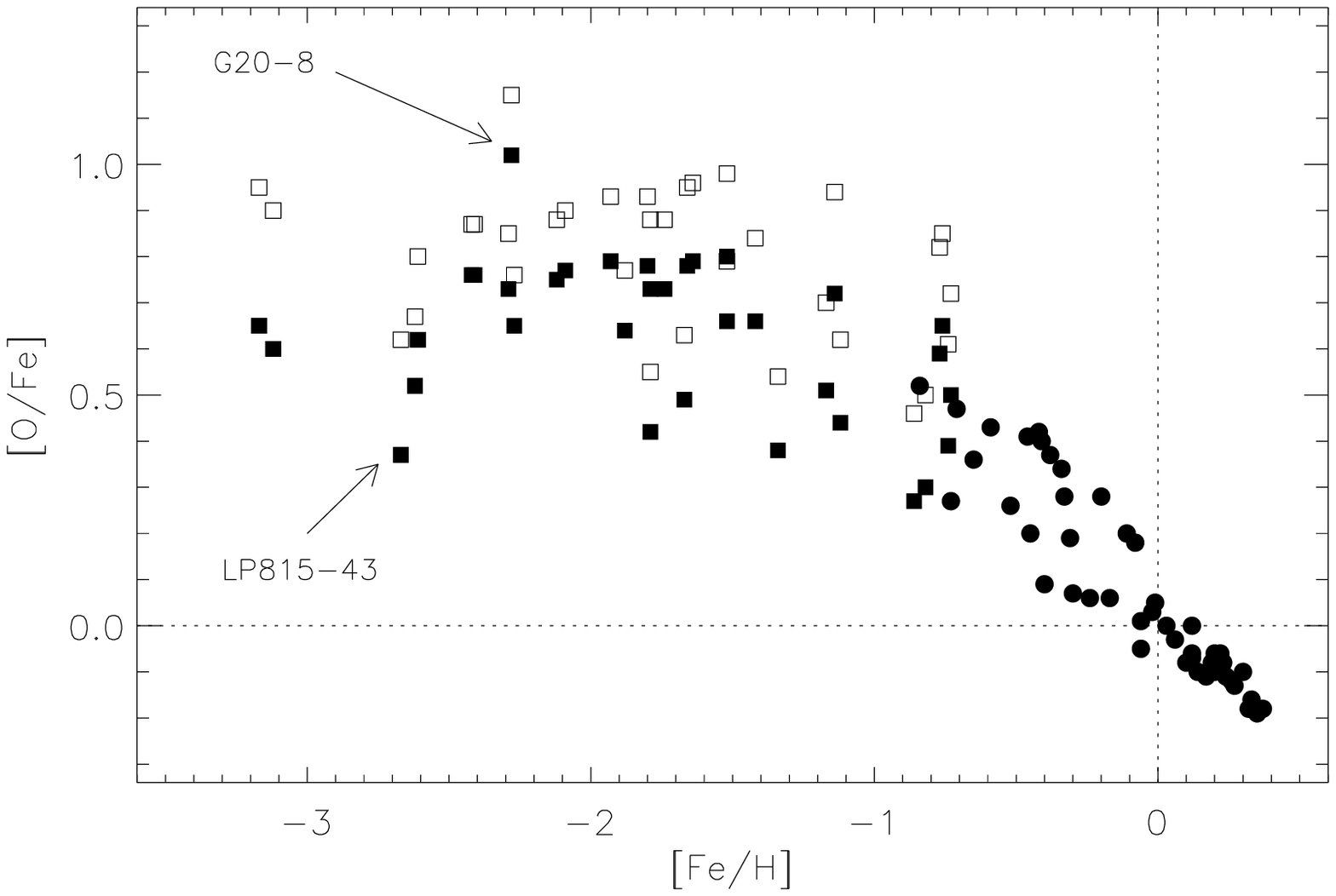}
\caption{Trends of [C/Fe] vs. \mbox{[Fe/H]}\, (left panel) and
[O/Fe] vs. \mbox{[Fe/H]}\,
 (right panel) in Milky Way halo (squares) and disk stars (circles,
 Bensby \& Feltzing 2006, from [C\,{\sc i}] and [O\,{\sc i}] lines free from non--LTE effects).
 The LTE values from Akerman et al. (2004),
 corrected for our different choice of carbon and oxygen solar
 abundances, are represented as empty squares in the two panels, while
 the non--LTE abundances are shown as filled squares. The non--LTE
 results are those obtained when neglecting collisions with
 hydrogen. Note that the [O/Fe] non--LTE values in the right plot for
 the five most metal-poor stars already include the new O\,{\sc i}\, non--LTE
 abundance corrections we present here (see text)}
\label{fabf:CandOovFe}
\end{center}
\end{figure*}

\end{flushleft}

\begin{figure*}[!ht]
\begin{center}
\includegraphics[width=12cm]{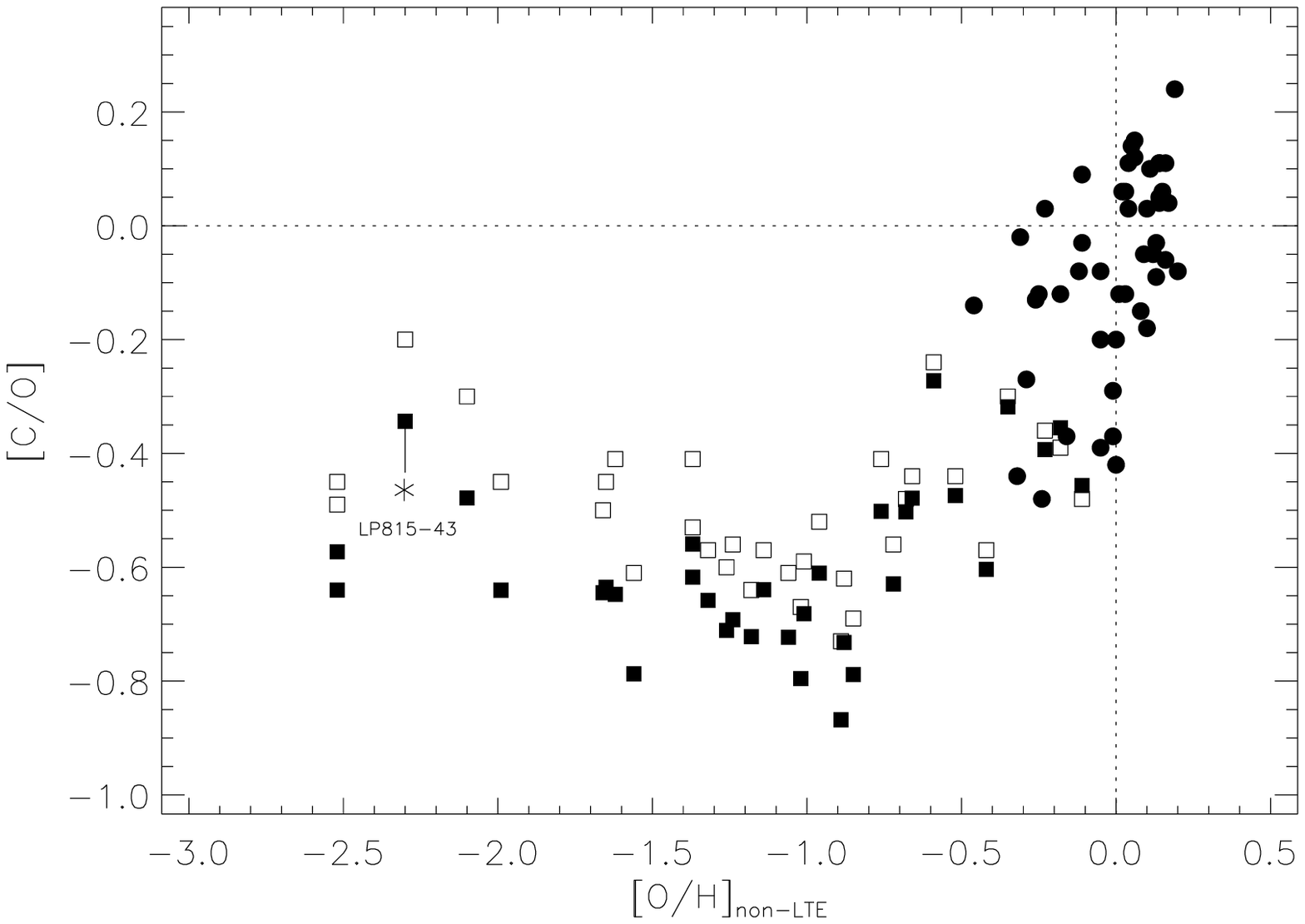}\\
\includegraphics[width=12cm]{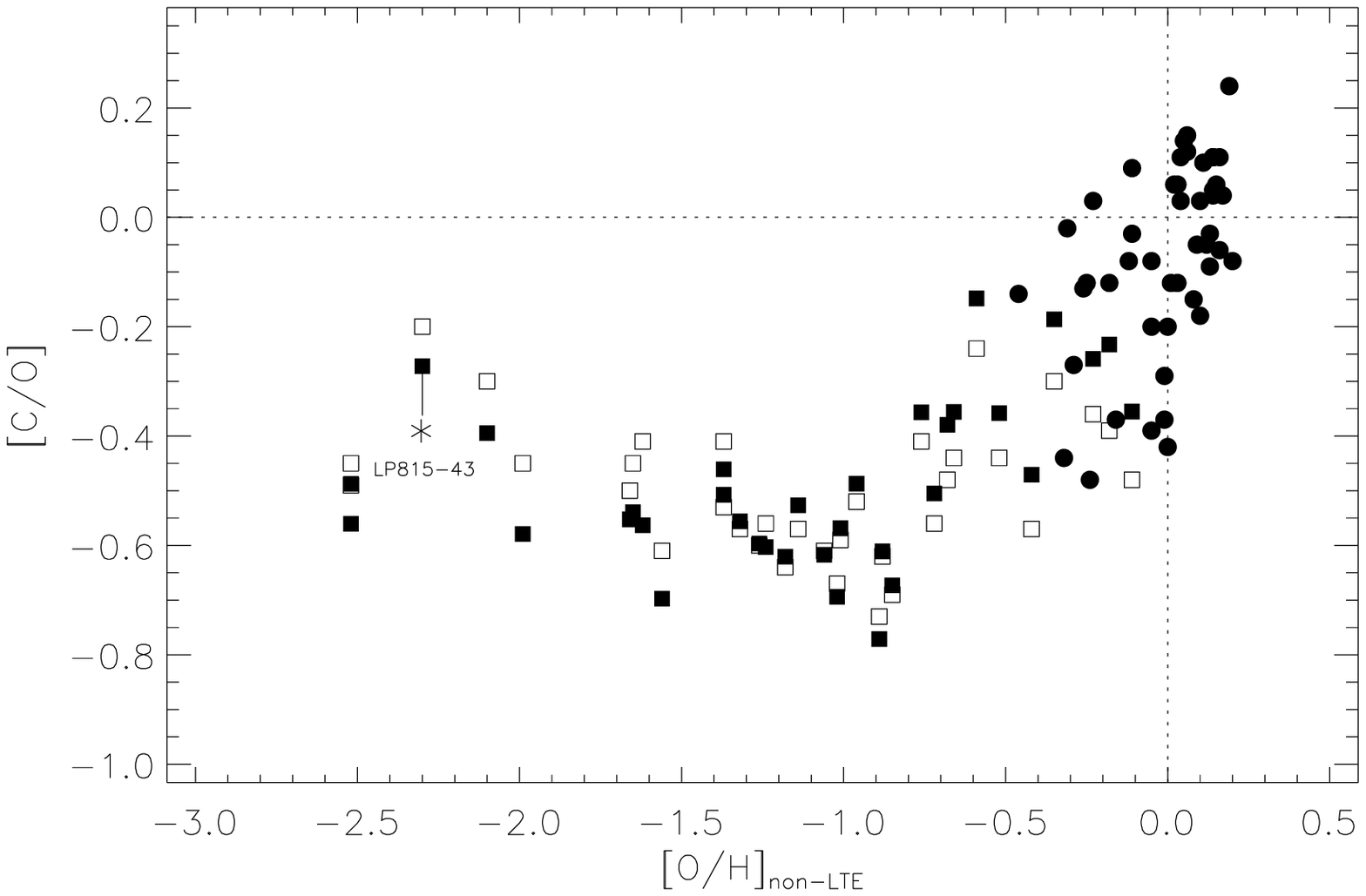}
 \caption{Trend of [C/O] vs. [O/H] in Milky Way halo and disk stars.
 The upper panel represents the new observational trends obtained when
 our non--LTE calculations are applied to the data in Akerman et
 al. 2004 (using our different choice of carbon and oxygen solar
 abundances and assuming negligible collisions with hydrogen). The
 [O/H] values already take into account our more negative non--LTE
 corrections for oxygen at low metallicity. The empty and filled
 squares represent the [C/O] LTE and non--LTE values respectively. The
 lower panel uses the same notation, with results now calculated
 accounting for hydrogen collisions for carbon (with a choice of
 S$_{\rm{H}}=1$). In both panels, the starred symbol indicates the [C/O]
 value we believe is more realistic for LP815-43 (see discussion in
 text)}
\label{fabf:GCE}
\end{center}
\end{figure*}

Equipped with our non--LTE calculations, we have revisited the
observational study of C and O in halo stars by Akerman et al.
(2004). Those authors found that [C/O] seems to rise for
\mbox{[O/H]}$<-1$; which could suggest high carbon yields from
zero-metallicity, so called Population III stars. They however assumed
that the non--LTE abundance corrections for the two elements would be
the same. We have calculated specific non--LTE corrections for their
sample of 34 dwarf halo stars. Since in their analysis those authors
used the same atmospheric models as we employ here, our derived
non--LTE corrections can be directly applied to their LTE abundance
results. Considering their estimate for random errors in equivalent
widths measurements and stellar parameters determination - which in
turn affects their resulting LTE abundances - and the uncertainty in
our non--LTE calculations ($\sim 0.1$~dex), we consider it feasible to
study the C and O trends. Moreover, the [C/O] ratio is essentially
insensitive to changes in the stellar parameters, given that the C and
O high-excitation abundance indicators used are affected in a similar
way. We are thus confident that our results are a step forward with
respect to previous works.

Table 2 lists the results obtained for the C and O abundances.
Fig.\,\ref{fabf:corrteff} shows the trend with atmospheric
parameters of the carbon non--LTE corrections we obtained for the
sample in Akerman et al. (2004). A clear increase in the non--LTE
abundance corrections (becoming approximately $0.3$~dex more
negative at \mbox{[Fe/H]}$\sim -3$ than at \mbox{[Fe/H]}$\sim -1$)
is seen as one moves to low metallicity, which is mostly due to the
higher effective temperatures of the sample stars towards the
metal-poor regime (\mbox{$T_{\rm{eff}}$}\, values increase by
approximately $500$~K or more for stars below \mbox{[Fe/H]}$\sim
-2$, compared to the higher-metallicity stars in the sample). This
is still in agreement with the fact that in general, when only
decreasing the metallicity of the atmospheric model to
\mbox{[Fe/H]}$\sim -3$, the non--LTE abundance corrections are
roughly of the same order than at solar metallicity (see
Figs.\,\ref{fabf:corrcontours} and \ref{fabf:splittedH0.001}). A
similar increase below \mbox{[Fe/H]}$\sim -1$ in the non--LTE
abundance corrections of Akerman et al.'s sample of halo stars was
found by Takeda \& Honda (2005), however when we adopt their same
choice for the H collision efficiency (S$_{\rm H}=1$) and
microturbulence ($\xi=1$~km/s), we still find corrections that are
more negative than theirs by $\sim 0.05-0.10$~dex for the
lowest-metallicity stars in that sample.

Previous studies of Milky Way disk stars (e.g. Gustafsson et al.
1999) found a slowly decreasing [C/Fe] with increasing metallicity
(slope of $-0.17$~dex per dex). However, very recently, Bensby \&
Feltzing (2006), analysing a sample of thin and thick disk stars and
using [C\,{\sc i}] and [O\,{\sc i}] lines immune from non--LTE
effects, found a more or less flat trend from around solar
metallicity down to \mbox{[Fe/H]}$\simeq -0.9$. We find a roughly
solar [C/Fe] (or slightly higher when adopting S$_{\rm H}=1$)
extending much further down to \mbox{[Fe/H]}$\simeq -3.2$
(Fig.\,\ref{fabf:CandOovFe}), albeit with a rather large and - at
least in part - likely real scatter. The plot shows how the LTE
[C/Fe] values in the Akerman et al. study have a large scatter
around a value of $\sim +0.35$~dex, with a possible hint for a
slight increase with decreasing \mbox{[Fe/H]}. This trend is removed
when correcting for our increasingly large (as one moves down in
metallicity) non--LTE corrections for the stars in their sample.  In
Fig.\,\ref{fabf:CandOovFe} we also show [O/Fe] as a function of
\mbox{[Fe/H]}\, for the halo stars of Akerman et al. As for the
[C/Fe] versus \mbox{[Fe/H]}\, plot, the disk stars of Bensby \&
Feltzing are included too. The original results of Akerman et al.
have here been corrected for our revised O\,{\sc i}\, non--LTE
corrections at the lowest \mbox{[Fe/H]}\, as well as for our
different solar abundances (Asplund et al. 2005b). The [O/Fe] trend
in metal-poor stars is still a hotly debated topic and a full
account of this problem is beyond the scope of the present paper; we
refer the reader to recent discussions such as Nissen et al. (2002)
and Asplund (2005) and references therein. We note however that
Fig.\,\ref{fabf:CandOovFe} suggests a roughly flat trend at a level
[O/Fe]$\sim +0.6$ with no indication of a linear trend with slope of
$\sim -0.3$~dex/dex as claimed by Israelian et al. (1998) and
Boesgaard et al. (1999). Asplund \& Garc\'{\i}a P{\'e}rez (2001)
have argued based on 3D hydrodynamical model atmospheres of
metal-poor stars that O abundances derived from OH lines as used by
Israelian et al. and Boesgaard et al. can be severely overestimated
in 1D (see also Collet et al. 2006).

A few peculiar cases in the two plots in Fig.\,\ref{fabf:CandOovFe}
should be noted:

\begin{itemize}

\item The C and O LTE abundances for \object{G16-13} (\mbox{[Fe/H]}$=-0.76$) are both
increased by $0.4$~dex with respect to Akerman et al. (2004), due to
the fact that a too hot stellar model was used in their analysis for
this star (P.E. Nissen, private communication). This moves the star
from the lower to the upper extreme of the distribution in the
[C/Fe] and [O/Fe] versus \mbox{[Fe/H]}\, plots.  Because the scatter
in both plots is rather large, this does not significantly affect
the overall resulting trends. Moreover, since both the C and O
abundances are increased by the same amount, the star maintains the
same [C/O] as in Akerman et al. (demonstrating how insensitive [C/O]
is to the model used), while [O/H] increases by $0.4$~dex: the
overall [C/O] trend is unchanged too, again because of masking by
the relatively large scatter at the higher-metallicity extreme of
the distribution, to which the star is shifted

\item The two most extreme cases in Fig.\,\ref{fabf:CandOovFe}
are \object{LP815-43} (with a low [O/Fe]$_{\rm non-LTE}=0.37$) and
\object{G20-8} (showing very high [C/Fe]$_{\rm non-LTE}=0.31$ and
[O/Fe]$_{\rm non-LTE}=1.02$). We note that for the O\,{\sc i}\,
triplet equivalent widths in LP815-43, Nissen et al. (2002) give
different values than those in Akerman et al. (2004). The smaller
line-to-line scatter in the former determination seems to better
correspond to the variation in gf-values; moreover, the S/N for this
star is higher in Nissen et al.'s spectrum. These authors derive
$\log\,\epsilon_{\rm O}=6.71$ compared to $\log\,\epsilon_{\rm
O}=6.61$ in Akerman et al. (2004). It is thus possible that the oxygen
abundance derived in the latter work needs to be increased by up to
$0.1$~dex, with the errors in the equivalent width determinations
likely attributable to e.g. continuum placement problems. The case of
G20-8, which shows anomalously large [C/Fe] and [O/Fe] values (by
$\sim 0.3$~dex), seems to indicate that its temperature might be
underestimated by as much as $400$~K, possibly due to uncertainties in
the reddening

\end{itemize}

The O\,{\sc i} non--LTE abundance corrections are taken from Akerman
et al. (2004), with the exception of the five most metal-poor stars.
Akerman et al. interpolated the non--LTE corrections of Nissen et
al. (2002) as a function of O\,{\sc i}\, equivalent width, leading
them to deduce a non--LTE correction of $-0.10$~dex for those five
stars. We have carried out non--LTE calculations in the same manner
as described by Nissen et al. Our corrections for the five most
metal-poor stars are more negative than found by Akerman et al.
(2004), varying between $-0.18$ and $-0.30$~dex. Moreover, while -
with respect to [C/O] - we find that the non--LTE corrections for
carbon and oxygen almost balance each other at higher metallicity,
at [O/H]$\apprle -1$ they become more negative for carbon. While the
O\,{\sc i}\, non--LTE corrections are typically $\simeq -0.2$~dex,
and $-0.3$~dex at most at \mbox{[Fe/H]}\,$\simeq -3$, the
corresponding C\,{\sc i}\, non--LTE corrections are $\simeq
0.15$~dex lower still at these low \mbox{[Fe/H]}. We adopt slightly
different solar C and O abundances than done by Akerman et
al.\footnote{Namely, $\log\epsilon_{C_\odot}=8.39$ and
$\log\epsilon_{O_\odot}=8.66$ (Asplund et al. 2004; Asplund et al.
2005b), as compared to $\log\epsilon_{C_\odot}=8.41$ and
$\log\epsilon_{O_\odot}=8.74$ in Akerman et al. (2004). We believe
the values we adopt are best estimates since they are averaged over
different lines and they account for 3D/non-LTE effects}, which also
slightly modifies the resulting [C/H], [O/H] and [C/O] values. Had
we used the solar values determined with a 1D MARCS model atmosphere
instead, [O/H] would have been lower by -0.05 dex and [C/H] higher
by +0.04 dex only, for all stars. Clearly, such a small and constant
offset is not in any way crucial in determining the general trends
of [C/Fe] and [O/Fe], because all stars would shift by the same
amount. More importantly, the [C/O] values remain essentialy
unchanged. Indeed, we recover a similar trend to that in the work by
Akerman et al. for the [C/O] abundance ratio at low metallicity,
even though at a lower level (see Fig.\,\ref{fabf:GCE}). We note
that the similarity is due to the fact that the carbon non--LTE
corrections in our study are more negative at low metallicity than
those we estimate for oxygen, by approximately the same amount for
all of the metal-poor stars below [O/H]$\sim -1.5$. This gives a
more or less constant downward shift of those [C/O] values as
compared to the Akerman et al. study, who had assumed for carbon
that the same non-LTE corrections would applicable as for oxygen.
The revised (more negative) O\,{\sc i}\, non--LTE abundance
corrections we use for the five most metal-poor stars in the sample
partially compensate the large non--LTE corrections we find for
C\,{\sc i} at such low metallicities. These increased non--LTE
corrections for oxygen also shift those stars to correspondingly
lower [O/H]$_{non-LTE}$ values with respect to the original study by
Akerman et al., ``stretching'' the low-metallicity part of the plot
with respect to their work.

In summary, the adoption of the recent values of solar abundance and
the application of our carbon and oxygen non--LTE corrections has the
net effect of shifting down the [C/O] values of the low-metallicity
stars by similar amounts, preserving the possible [C/O] upturn at
[O/H]$\apprle -1$, although at a lower [C/O] level. We do urge that
this rising trend should not be over-interpreted at this stage.  First
of all, if our suspicion is confirmed that the oxygen abundance has
been underestimated by $\sim 0.1$~dex by Akerman et al. for LP815-43 -
which is the star with the highest [C/O] ratio at low metallicity - it
would imply that this star will accordingly have a lower [C/O],
bringing it closer to the rest of the low-metallicity points and
giving in the end a less significant indication for the claimed
upturn. The [C/O] trend below [O/H]$\sim -1$ might thus have a gradual
increase with a shallow slope of $\sim -0.2$~dex/dex.  Moreover, in
view of the large departures from LTE we found for C\,{\sc i}\, in 1D, it is
important to carry out 3D non--LTE calculations. Furthermore, our
tests have revealed that the O\,{\sc i}\, non--LTE effects at these very low
\mbox{[Fe/H]}\, may be more uncertain than previously thought due to large
sensitivity to poorly known collisional cross-sections. We are indeed
planning to investigate this in an upcoming work. Finally, the
\mbox{$T_{\rm{eff}}$}\, scale for halo stars is still under debate (e.g. Mel\'{e}ndez et
al. 2006). The claim of carbon production from high-mass Pop. III
stars is thus still uncertain in our opinion.

Spite et al. (2005) used different abundance indicators (namely, the
G band of CH and the forbidden [O\,{\sc i}] line at 630.0 nm) to
extend the study by Akerman et al. down to \mbox{[Fe/H]}\,$\sim -4$
(\mbox{[O/H]}\,$\sim -3$), again deriving an increasing [C/O] ratio
reaching a near solar level ($\sim -0.2$ dex) toward lower
metallicity. However, they did not account for 3D effects which
should affect the carbon molecular lines they employed more than the
[O\,{\sc i}] line they used. Even though those authors studied
extremely metal-poor halo giants, for which 3D effects are not well
known, one can use the results for turn-off stars at similarly low
metallicity as an indication, so that the corresponding values given
in Spite et al. might turn out to be overestimated by $\sim 0.3$~dex
or more (Asplund 2005) and the [C/O] trend could still be roughly
constant or only slowly increasing.

\section{Conclusions}

We have carried out non--LTE spectral-line formation calculations for
C\,{\sc i} absorption features. The computations have been performed
on a grid of 168 1D {\small MARCS} stellar model atmospheres with
varying parameters ($4500\le$\mbox{$T_{\rm{eff}}$}$\le7000$~K, $2\le$\mbox{log $g$}$\le5$,
$-3\le$\mbox{[Fe/H]}$\le0$) and $-0.60\le$[C/Fe]$\le+0.60$). The main results
are:

\begin{itemize}

\item We generally find large, negative non--LTE abundance corrections
for high-excitation neutral carbon permitted lines, i.e. LTE
abundances are too large. At solar metallicity, the dominant non--LTE
effect is that of the line source function dropping below the
Planckian mostly due to under-populated upper atomic levels in the
transitions, but at low metallicity it is that of line opacity due to
overpopulated lower levels

\item We find substantial non--LTE abundance corrections for metal-poor
halo turn-off stars: typically $\sim -0.4$~dex at \mbox{[Fe/H]}$\sim -3$; these
results are relatively immune to the adopted H collisional efficiency

\item Using our resulting non--LTE corrections, we revisit the observational
study of Akerman et al. (2004) and find that the [C/O] trend at low
metallicity might still be suggesting an upturn indicating possible
production of carbon through nucleosynthesis in Pop. III stars. We
however warn that more observational data is needed - together with
detailed non--LTE calculations for oxygen, to understand the magnitude
of the abundance corrections applicable for this element at
metallicities \mbox{[Fe/H]}$\apprle-2.5$ - in order to reduce the still large
uncertainties and finally settle this issue

\end{itemize}

Without doubt one of the necessary steps for obtaining reliable
stellar abundances in future large scale surveys will be to perform a
vast quantity of detailed non--LTE studies for an increased number of
different chemical elements and stellar parameters. Very many precise
atomic input data for collisional and radiative processes have to be
known for realistic results. The use of sophisticated techniques for
quantum-mechanical computations have led to accurate atomic radiative
data fortunately becoming available in recent years to the
astrophysical community, thanks to such large scale efforts as e.g. the
Opacity and IRON Projects. We underline the necessity for more
theoretical and laboratory studies aimed at obtaining such accurate
atomic data and in particular at improving our knowledge on the
correct treatment of the still poorly understood collisions with
electrons and with hydrogen atoms. The effect of using 3D
hydrodynamical model atmospheres should be investigated as the 3D
non--LTE effects could prove to be even stronger than in 1D, due to
the presence of large temperature inhomogeneities and to the different
temperature structure in the optically thin atmospheric layers with
respect to 1D hydrostatic models, especially for the low-metallicity
regime (Asplund et al. 1999).

\bigskip
\begin{acknowledgements}
DF acknowledges support from RSAA, in the form of a Joan Duffield
Research Scholarship and an Alex Rodgers Travelling Scholarship and is
grateful to the Institute of Theoretical Astrophysics, University of
Oslo, Norway, for its repeated hospitality. DF also thanks \O ystein
Langangen and Jorge Mel\'{e}ndez for useful discussions. This work has
been partly funded by the Australian Research Council (grants
DP0342613 and DP0558836) and also supported by the Research Council of
Norway through grant 146467/420 and through a grant of computing time
from the Programme for Supercomputing. This research has made
extensive use of NASA's Astrophysics Data System and of the NIST
Atomic Spectra Database, which is operated by the National Institute
of Standards and Technology. Finally, we would like to thank the
referee for remarks that were useful in improving the paper.
\end{acknowledgements}


\begin{thebibliography}{}

\bibitem[]{} Akerman, C.~J., Carigi, L., Nissen, P.~E. et al. 2004, A\&A, 414, 931
\bibitem[]{} Allende Prieto, C., Asplund, M., L\'{o}pez, Ram\'{o}n J.~G., Lambert, D.~L. 2002, ApJ, 567, 544
\bibitem[]{} Andersson, H., \& Edvarsson, B. 1994, A\&A, 290, 590
\bibitem[]{} Arnett, D. 1996, {\it Supernovae and Nucleosynthesis} (Princeton University, Princeton, NJ)
\bibitem[]{} Asplund, M., Gustafsson, B., Kiselman, D., \& Eriksson, K. 1997, A\&A, 318, 521
\bibitem[]{} Asplund, M.,  Nordlund, \AA, Trampedach, R., \& Stein, R.~F. 1999, A\&A,
346, L17
\bibitem[]{} Asplund, M., \& Garc\'{\i}a P{\'e}rez A.~E. 2001, A\&A, 372, 601
\bibitem[]{} Asplund, M., Grevesse, N., Sauval, A.~J. et al. 2004, A\&A, 417, 751
\bibitem[]{} Asplund, M. 2005, ARA\&A, 43, 481   %
\bibitem[]{} Asplund, M., Grevesse, N., Sauval, A.~J. et al. 2005a, A\&A, 431, 693   %
\bibitem[]{} Asplund, M., Grevesse, N., \& Sauval, A.~J. 2005b, ASP Conference Series, 336, 25   %
\bibitem[]{} Ballero, S.~K, Matteucci, F., \& Chiappini, C. 2006, New Astronomy, 11, 306
\bibitem[]{} Becker, G., D., Sargent W.~L.~W., Rauch, M., \& Simcoe, R.~A. 2006, ApJ, 640, 69
\bibitem[]{} Belyaev, A.~K., Grosser, J., Hahne, J., \& Menzel, T. 1999, Phys. Rev. A, 60, 2151
\bibitem[]{} Belyaev, A.~K., \& Barklem, P.~S. 2003, Phys. Rev. A, 68, 62703
\bibitem[]{} Bensby, T., \& Feltzing, S. 2006, MNRAS, 367, 1181
\bibitem[]{} Boesgaard, A.~M., King, J.~R., Deliyannis, C.~P., Vogt, S.~S. 1999, AJ, 117, 492
\bibitem[]{} Burbidge, E.~M., Burbidge, G.~R., Fowler, W.~A., \& Hoyle, F. 1957, Rev. Mod. Phys., 29, 547
\bibitem[]{} Carigi, L., Peimbert, M., Esteban, C., \& Garc\'{\i}a-Rojas, J. 2005, ApJ, 623, 213
\bibitem[]{} Carlsson, M. 1986, Uppsala Astronomical Observatory Report No. 33
\bibitem[]{} Carlsson, M. 1992, ASP Conference Series, 26, 499
\bibitem[]{} Carlsson, M., Rutten, R.~J., Bruls, J.~H.~M.~J., \& Shchukina, N.~G. 1994, A\&A, 288, 860
\bibitem[]{} Chiappini, C., Matteucci, F., \& Gratton, R. 1997, ApJ, 477, 765
\bibitem[]{} Chiappini, C., Romano, D., \& Matteucci, F. 2003, MNRAS, 339, 63
\bibitem[]{} Chiappini, C., Hirschi, R., Meynet, G. et al. 2006, A\&A, 449, L27
\bibitem[]{} Collet, R., Asplund, M., Trampedach, R. 2006, ApJL accepted (astro-ph/0605219)
\bibitem[]{} Cunto, W., Mendoza, C., Ochsenbein, F., \& Zeippen, C. 1993, A\&A, 275, L5
\bibitem[]{} Dopita, M. A., \& Sutherland, R. S. 2003, Astrophysics of the diffuse universe, Springer, 2003
\bibitem[]{} Drawin, H.~W. 1968, Zeitschrift f. Physik, 211, 404
\bibitem[]{} Erni, P., Richter, P., Ledoux, C., \& Petitjean, P. 2006, A\&A, 451, 19
\bibitem[]{} Fabbian, D., Asplund, M., Carlsson, M., \& Kiselman, D. 2005, in IAU Symp. 228, From Lithium to Uranium: Elemental Tracers of Early Cosmic Evolution, Hill, Fran\c{c}ois \& Primas eds., Cambridge University
Press, p. 255
\bibitem[]{} Fleck, I., Grosser, J., Schnecke, A. et al. 1991, J. Phys. B, 24, 4017
\bibitem[]{} Gavil\'{a}n, M., Buell, J.~F. \& Moll\'{a}, M.  2005, A\&A, 432, 861
\bibitem[]{} Gray, D.~F. 1992, The Observation and Analysis of Stellar Photospheres, $2^{nd}$ ed., Cambridge University Press
\bibitem[]{} Gustafsson, B., Bell, R.~A., Eriksson, K., \& Nordlund, \AA. 1975, A\&A, 42, 407
\bibitem[]{} Gustafsson, B., Karlsson, T., Olsson, E. et al. 1999, A\&A, 342, 426
\bibitem[]{} Hibbert, A., Biemont, E., Godefroid, M., Vaeck, N. 1993, A\&A Supplement Series, 99, 179
\bibitem[]{} Israelian, G., Garc\'{\i}a L{\'o}pez, R.~J., \&  Rebolo, R. 1998, ApJ, 507, 805
\bibitem[]{} Kaulakys, B. 1985, J. Phys. B, 18, L167
\bibitem[]{} Levshakov, S.~A., Centuri\'{o}n, M., Molaro, P., \& Kostina, M.~V. 2006, A\&A, 447, L21
\bibitem[]{} Luo, D., \& Pradhan, A. K. 1989, Journal of Physics B, 22, 3377
\bibitem[]{} Magain, P. 1986, A\&A, 163, 135
\bibitem[]{} Mel\'{e}ndez, J., Shchukina, N.~G., Vasiljeva, I., \& Ram\'{i}rez, I. 2006, ApJ, 642, 1082
\bibitem[]{} Meynet, G., Ekstr\"{o}m, S., \& Maeder, A. 2006, A\&A, 447, 623
\bibitem[]{} Martin, W.~C., Dalton, G.~R., Fuhr, J.~R. et al. 1995, ASP Conference Series, 81, 597
\bibitem[]{} Mihalas, D. \& Mihalas, B.~W. 2000, Foundations of Radiation Hydrodynamics, Dover Publications
\bibitem[]{} Nissen, P.~E., Primas, F., Asplund, M., \& Lambert, D.~L. 2002, A\&A, 390, 235
\bibitem[]{} Pagel, B.~E.~J. 1997, {\it Nucleosynthesis and Chemical Evolution of Galaxies} (Cambridge University Press, Cambridge, England)
\bibitem[]{} Paunzen, E., Kamp, I., Iliev, I.~Kh. et al. 1999, A\&A, 345, 597
\bibitem[]{} Reddy, B.~E., Tomkin, J., Lambert, D.~L., Allende Prieto, C. 2003, MNRAS, 340, 304
\bibitem[]{} Reddy, B.~E., Lambert, D.~L., \& Allende Prieto, C. 2006, MNRAS, 367, 1329
\bibitem[]{} Rentzsch-Holm, I. 1996a, ASP Conference Series, 108, 99, Model Atmospheres and Spectrum Synthesis, Adelman, Kupka \& Weiss eds.
\bibitem[]{} Rentzsch-Holm, I. 1996b, A\&A, 312, 966
\bibitem[]{} Rutten, R.~J. 2003, Radiative Transfer in Stellar Atmospheres (http://www.phys.uu.nl/$\sim$rutten/Astronomy$\_$lecture.html)
\bibitem[]{} Spite, M., Cayrel, R., Plez, B. et al. 2005, A\&A, 430, 655
\bibitem[]{} Steenbock, W., \& Holweger, H. 1984, A\&A, 130, 319
\bibitem[]{} St\"{u}renburg, S., \& Holweger, H. 1990, A\&A, 237, 125
\bibitem[]{} Takeda, Y. 1992, PASJ, 44, 649
\bibitem[]{} Takeda, Y. 1994, PASJ, 46, 53
\bibitem[]{} Takeda, Y., \& Honda, S. 2005, PASJ, 57, 65
\bibitem[]{} Tomkin J., Lemke, M., Lambert, D.~L., \& Sneden, C. 1992, AJ, 104, 1568
\bibitem[]{} van Regemorter, H. 1962, ApJ, 136, 906
\bibitem[]{} Wiese, W.~L., Fuhr, J.~R., \& Deters, T.~M. 1996, J. Phys. Chem. Ref. Data Monograph, No. 7
\bibitem[]{ww95} Woosley, S.~E., \& Weaver, T.~A. 1995, ApJS, 101, 181
\end{thebibliography}
\end{document}